\documentclass[11pt]{article}
\usepackage{graphics}

\newcommand{\be}{\begin{equation}}
\newcommand{\ee}{\end{equation}}
\newcommand{\bea}{\begin{eqnarray}}
\newcommand{\eea}{\end{eqnarray}}

\newcommand{\slp}{\sin^2\phi}
\newcommand{\srp}{\sin^2\phi_{f,\,right}}
\newcommand{\slps}{\sin\phi}

\newcommand{\lae}{\stackrel{<}{\sim}}
\newcommand{\gae}{\stackrel{>}{\sim}}
\newcommand{\sm}{\hbox{\bf\sf m}}

\begin{document}
\begin{titlepage}
\def\thepage {}        

\title{Weak-Singlet Fermions: Models and Constraints}

\author{
Marko B. Popovic and Elizabeth H. Simmons\thanks{e-mail addresses: 
markopop@buphy.bu.edu, simmons@bu.edu},\\
Department of Physics, Boston University, \\
590 Commonwealth Ave., Boston MA  02215}

\date{\today}

\maketitle

\bigskip
\begin{picture}(0,0)(0,0)
\put(295,240){BUHEP-99-29}
\put(295,225){hep-ph/0001302}
\end{picture}
\vspace{24pt}

\begin{abstract}
  
  We employ data from precision electroweak tests and collider
  searches to derive constraints on the possibility that weak-singlet
  fermions mix with the ordinary Standard Model fermions.  Our
  findings are presented within the context of a theory with
  weak-singlet partners for all ordinary fermions and theories in
  which only third-generation fermions mix with weak singlets.  In
  addition, we indicate how our results can be applied more
  widely in theories containing exotic fermions.

\pagestyle{empty}
\end{abstract}
\end{titlepage}



\section{Introduction}
\label{sec:intro}
\setcounter{equation}{0}

The origins of electroweak and flavor symmetry breaking remain
unknown.  The Standard Model of particle physics describes both
symmetry breakings in terms of the Higgs boson.  Electroweak symmetry
breaking occurs when the Higgs spontaneously acquires a non-zero
vacuum expectation value; flavor symmetry breaking is implicit in the
non-universal couplings of the Higgs to the fermions.  However, the
gauge hierarchy and triviality problems imply that the Standard Model
is only an effective field theory, valid below some finite momentum
cutoff.  The true dynamics responsible for the origin of mass must
therefore involve physics beyond the Standard Model.  This raises the
question of whether the two symmetry breakings might be driven by
different mechanisms.  Many theories of non-Standard physics invoke
separate origins for electroweak and flavor symmetry breaking, and
place flavor physics at higher energies in order to satisfy
constraints from precision electroweak test and flavor-changing
neutral currents.

In this paper, we explore the possibility that flavor symmetry
breaking and fermion masses may be connected with the presence of
weak-singlet fermions mixing with the ordinary Standard Model
fermions.  Specifically, we consider theories in which some of the
observed fermions' masses arise through a seesaw mechanism that
results in the presence of two mass eigenstates for each affected
flavor: a lighter mass eigensate whose left-handed component is
predominantly weak-doublet, and a heavier one that is mostly
weak-singlet.  Such seesaw mass structures involving either
third-generation fermions \cite{tsee,them} or all fermions
\cite{burdev} have played a prominent role in recent work on dynamical
symmetry breaking.

This work uses published experimental data to elicit constraints on
the masses and mixing strengths of the exotic fermions.  We both
interpret our findings within the context of several specific models
and indicate where our results can be applied more widely.  Our
initial approach is to study Z-pole and low-energy data for signs that
the known fermions include a non-Standard, weak-singlet component.
Previous limits of this type \cite{langlon, nardi, bbp} have found
that the mixing fraction $\sin^2 \theta_{mix}$ can be at most a few
percent for any given fermion species.  As a complementary test we
also look for evidence that new heavy fermions with a large
weak-singlet component are being pair-produced in high-energy collider
experiments.  This can provide a direct lower bound on the mass of the
new fermions.  Most recent limits on production of new fermions focus
on sequential fermions (LH doublets and RH singlets), mirror fermions
(RH doublets and LH singlets), and vector fermions (LH and RH
doublets) \cite{PDG}. These limits need not apply directly to weak
singlet fermions, as their production cross-sections and decay paths
can differ significantly from those of the other types of fermions.

We take as our benchmark a model \cite{origpaper} in which each
ordinary fermion flavor mixes with a separate weak-singlet fermion;
this allows us to consider the diverse phenomenological consequences
of the singlet partners for quarks and leptons of each generation.
The low-energy spectrum is completely specified, so that it is
possible to calculate branching ratios and precision effects.
Electroweak symmetry breaking is caused by a scalar, $\Phi$, with
flavor-symmetric couplings to the fermions.  Flavor symmetry breaking
arises from physics at higher scales that manifests itself at low
energies in the form of soft symmetry-breaking mass terms linking
ordinary and weak-singlet fermions.  The fermions' chiral symmetries
enforce a GIM mechanism and ensure that the flavor structure is
preserved under renormalization.  Due to recent interest in using weak
singlets to explain the mass of the top quark \cite{tsee}, we also
analyze variants of our benchmark model in which only the
third-generation fermions have weak-singlet partners.  Furthermore, we
indicate how our results can be applied to other theories with
weak-singlet fermions.

Since our benchmark model includes a scalar boson, it should be
considered as the low-energy effective theory of a more complete
dynamical model; specifically, at some finite energy scale, the scalar
$\Phi$, like the Higgs boson of the Standard Model, would reveal
itself to be composite.  That more complete model would be akin to
dynamical top seesaw models \cite{tsee, them, burdev}, which include
composite scalars, formed by new strong interactions among quarks, and
also have top and bottom quarks' masses created or enhanced by mixing
with weak-singlet states.  Those particular top seesaw models
generally have multiple composite scalars when more than one fermion
has a weak-singlet partner; these tend to be heavier than the single
scalar in our models.  Moreover, in the ``flavor-universal'' versions
\cite{burdev} generation-symmetry-breaking masses for the weak singlet
fermions are the source of the differences between the masses of, say,
the up and top quarks; the flavor structure of our models is
different.  Despite these differences, most of our phenomenological
results are relevant to the top seesaw models.

In the next section, we review the structure of our benchmark model, focusing
on the masses, mixings, and couplings of the fermions.  Section 3 discusses
our fit to precision electroweak data \cite{LEPEWWG} and the resulting
general limits on the mixing angles between ordinary and weak-singlet
fermions.  We then use the constraints on mixing angles
to find lower bounds on the masses of the new heavy fermion eigenstates.
Section 5 discusses the new fermions' decay modes and extracts lower bounds
on the fermion masses from LEP II \cite{OPAL-lepton, DELPHI-lepton} and Tevatron
\cite{D0-quark, CDF-quark} data.  Oblique corrections are discussed in
section 6 and our conclusions are summarized in the final section.

\section{The Model}
\label{sec:model}
\setcounter{equation}{0} 

At experimentally accessible energies, the models we consider have the
gauge group of the Standard Model: $SU(3)_C \times SU(2)_W \times
U(1)_Y$.  The gauge eigenstate fermions include three generations of ordinary
quarks and leptons, which are left-handed weak doublets and right-handed
weak singlets
\bea 
\psi_{L} &=& {U \choose D}_{L} ,\ \ \ \ U_{R},\ \ D_{R}\ \ \ \ \ \ \ 
U \equiv (u, c, t) ,\ \ D \equiv (d, s, b)\ ,\nonumber \\
L_L &=& {\nu_\ell \choose \ell}_L ,\ \ \ \ \ \ell_R\ \ \ \ \ \ \ \ \ \ \ \
\ \ 
\ \
\ell \equiv (e, \mu, \tau)\ .
\label{ordferm}
\eea
In our general, benchmark model to each `ordinary' charged fermion there
corresponds a `primed' weak-singlet fermion with the same electric
charge\footnote{In principle, one could include weak singlet partners for the
  neutrinos as well.  The neutrino phenomenology is largely separate from the
  issues treated here and will not be considered in this paper.}
\be
U^\prime_{L,R},\ \ D^\prime_{L,R},
                      \ \ \ell^\prime_{L,R}\  .
\label{heavy}
\ee
We will also discuss the phenomenology of more specialized models in which only
third-generation fermions have `primed' weak-singlet partners.

The gauge symmetry allows bare mass terms for the weak-singlet fermions
\be
 M_U \bar U^\prime_L U^\prime_R + M_D \bar
  D^\prime_L D^\prime_R + M_\ell \bar \ell^\prime_L \ell^\prime_R
\label{baremass}
\ee
and we take each of these mass matrices $M_f$ to be proportional to
the identity matrix.

The model includes a scalar doublet field
\be
\Phi = {\Phi^+ \choose \Phi^0}
\label{higgs}
\ee
whose VEV breaks the electroweak symmetry.  This scalar has 
Yukawa couplings that link left-handed ordinary to right-handed primed
fermionic gauge eigenstates
\be
\lambda_U \bar\psi_L \tilde \Phi\thinspace  U^\prime_R + 
            \lambda_D \bar\psi_L \Phi\thinspace  D^\prime_R +
            \lambda_\ell \bar L_L \Phi\thinspace \ell^\prime_R \ .
\label{yukawa}
\ee
The coupling matrices $\lambda_f$ are taken to be proportional to the
identity matrix.  The mass of the scalar is assumed to be small enough that
the scalar's contributions will prevent unitarity violation in scattering of
longitudinal weak vector bosons.

Finally, there are mass terms connecting left-handed primed and right-handed
ordinary fermions
\be
 \bar U^\prime_L \sm_U U_R + \bar D^\prime_L
  \sm_D D_R + \bar \ell^\prime_L \sm_\ell \ell_R .
\label{fsbmass}
\ee
which break the fermions' flavor symmetries.  We shall require the
flavor-symmetry violation to be small: any mass $\sm_f$ should be no greater
than the corresponding mass $M_f$.  This allows our model to
incorporate the wide range of observed fermion masses without jeopardizing
universality \cite{origpaper}.

As discussed in reference \cite{origpaper}, this flavor structure is stable under
renormalization.  On the one hand, the flavor-symmetry-breaking mass terms
(\ref{fsbmass}) are dimension-three and cannot renormalize the
flavor-symmetric dimension-four Yukawa terms (\ref{yukawa}).  On the other,
because all dimension-four terms (including the Yukawa couplings
(\ref{yukawa})) respect the full set of global chiral symmetries,
\bea
SU(3)_{\psi_L,U^\prime_R,D^\prime_R}\!\!&\times& SU(3)_{U^\prime_L}\times
SU(3)_{D^\prime_L}\times SU(3)_{U_R}\times SU(3)_{D_R}\times \nonumber \\
&\,&SU(3)_{L_L, l^\prime_R}\times SU(3)_{l^\prime_L}\times
SU(3)_{l_R} 
\eea
they do not mix the mass terms (\ref{baremass}) and (\ref{fsbmass}) which
break those symmetries differently.  Furthermore, the global symmetries of
this model lead to a viable pattern of inter-generational mixing among the
fermions.  Including the $M_f$ terms (\ref{baremass}) breaks the flavor
symmetries to a form
\be
SU(3)_{\psi_L,U^\prime,D^\prime}\times SU(3)_{U_R}\times SU(3)_{D_R}
\times SU(3)_{L_L,l^\prime}\times SU(3)_{l_R}
\ee
nearly identical to that of the Standard Model with massless fermions.  Once
the flavor-symmetry-breaking masses of equation (\ref{fsbmass}) are added,
the quarks' flavor symmetries are completely broken, leading to the presence
of a CKM-type quark mixing matrix and an associated GIM mechanism that
suppresses flavor-changing neutral currents.  The lepton sector retains the
$U(1)$'s corresponding to conservation of three separate lepton numbers.

\bigskip

The ordinary and primed fermions mix to form mass eigenstates; for each type
of charged fermion ($f \equiv $ $U$, $D$, $\ell$) the mass matrix in the
gauge basis is of the form
\be
\left(\bar{f}\ \bar{f}^\prime\right)_L\ \ 
\left( \matrix{0 & v \lambda_f \cr
                            \sm_f  & M_f\cr } \right)
\ {f \choose f^\prime}_R\ .
\label{massmat}
\ee
This is diagonalized by performing separate rotations on the left-handed and
right-handed fermion fields. The phenomenological issues we shall examine
will depend almost exclusively on the mixing among the left-handed fermions.
Hence, our discussion related to fermion mixing and its effects will focus on
the left-handed fermion fields.  For brevity, we omit ``left'' subscripts 
on the left-handed mixing angles and fields; we include ``right'' subscripts
in the few instances where the right-handed mixings play a role.

To evaluate the degree of mixing among the left-handed weak-doublet and
weak-singlet fields, we diagonalize the mass-squared matrix $(M^\dagger
M)$.  The rotation angle among left-handed fermions is given by\footnote{To
  study the rotation angle among right-handed fermions, one diagonalizes $(M
  M^\dagger)$ and obtains analogous results.}
\bea
\sin^2\phi_f = 1 - {B^2 \over{ 2 A^2 + 2 B^2 - 2 A \sqrt{A^2 + B^2}}}
\ \ \ \ \ \ \ \ \ \ B \!\!&=&\!\! 2 v \lambda_f M_f \label{sineq}\\
 A \!\!&=&\!\! M_f^2 + \sm_f^2 - v^2 \lambda_f^2\nonumber
\label{analggy}
\eea
and the mass-squared eigenvalues are
\be
\Lambda_f^\pm = {1\over2}
\left(M_f^2 + \sm_f^2 + v^2 \lambda_f^2 \right)
\left( 1 \pm \sqrt{1 - (4 v^2 \lambda_f^2 \sm_f^2)/
                         (M_f^2 + \sm_f^2 + v^2 \lambda_f^2)^2}\right) .
\label{lameq}
\ee
Due to the matrix's seesaw structure, one mass eigenstate
($f^L$) has a relatively small mass, while the mass of the other eigenstate
($f^H$) is far larger.  The lighter eigenstate, which has a left-handed
component dominated by the ordinary weak-doublet state,
\be
f^L = \cos\phi_f f - \sin\phi_f f^\prime\ ,
\label{liferm}
\ee
corresponds to one of the fermions already observed by experiment.  Its mass
is approximately given by (for $ v \lambda_f < M_f$ and $\sm_f \le M_f$)
\be
(m_f^L)^2 \approx {(v \lambda_f \sm_f)^2 \over
M_f^2 + (v \lambda_f)^2 + \sm_f^2}\ .
\label{limass}
\ee
The heavier eigenstate, whose left-handed component is largely
weak-singlet,
\be
f^H = \sin\phi_f f + \cos\phi_f f^\prime
\label{heferm}
\ee
has a mass of order
\be
(m_f^H)^2 \approx M_f^2 + (v \lambda_f)^2 + \sm_f^2
- (m_f^L)^2\ . 
\label{hemass}
\ee

The interactions of the mass eigenstates with the weak gauge bosons differ
from those in the Standard Model because the primed fermions lack weak
charge\footnote{For a general discussion of fermion mixing and gauge
  couplings in the presence of exotic fermions, see \cite{langlon}.}.  The
coupling of $f^L$ ($f^H$) to the W boson is proportional to 
$\cos\phi_f$ ($\sin\phi_f)$; the right-handed states are purely weak-singlet
and do not couple to the $W$ boson.  Thus the couplings of left-handed
leptons to the $W$ boson look like (since we neglect neutrino mixing)
\be
{\frac{ie}{\sin\theta_W}}\, \left( \ell^L \gamma_\mu \bar{\nu_\ell}
  \cos{\phi_\ell} +  
\ell^H \gamma_\mu \bar{\nu_\ell} \sin{\phi_\ell} \right) W^\mu
\label{lwco}
\ee
When weak-singlet partners exist for all three generations of quarks, the
left-handed quarks' coupling to the $W$ bosons is of the form
\be
     {\frac{ie}{\sin\theta_W}}\, (\bar{U^L}, \bar{U^H})\gamma_\mu 
     V_{UD} \left 
     ( \begin{array}{c} D^L \\ 
        D ^H \end{array} \right) W^\mu 
\ee
The $6 \times 6$ non-unitary matrix $V_{UD}$ is related to the underlying
$3\times 3$ unitary matrix $A_{UD}$ that mixes quarks of different
generations
\be
    V_{UD} = \left( \begin{array}{cc}
      C_U\  A_{UD}\  C_D\ \  &\ \  - C_U\  A_{UD}\  S_D \\
    - S_U\  A_{UD}\ C_D\ \  & \ \   S_U\ A_{UD}\ S_D \\
      \end{array} \right)\label{Vudu}
\ee
through diagonal matrices of mixing factors
\bea
C_U \equiv diag(\cos{\phi_u},\, \cos{\phi_c},\, \cos{\phi_t})\,, \ \ &\ &\ \ 
C_D \equiv diag(\cos{\phi_d},\, \cos{\phi_s},\, \cos{\phi_b})\,, \nonumber \\
S_U \equiv diag(\sin{\phi_u},\, \sin{\phi_c},\, \sin{\phi_t})\,, \ \ &\ &\ \ 
S_D \equiv diag(\sin{\phi_d},\, \sin{\phi_s},\, \sin{\phi_b})\,. \nonumber 
\eea
The unitary mixing matrix $A_{UD}$, like the CKM matrix in the Standard
Model, is characterized by three real angles and one CP-violating phase.  But it
is the elements of $V_{UD}$ which are directly accessible to experiment.
While $V_{UD}$ is non-unitary, any two columns (or rows) are still orthogonal.
        
The coupling of left-handed mass-eigenstate fermions to the $Z$ boson
        is of the form 
\be
{\frac{ie}{\sin\theta_W\cos\theta_W}}\, (\bar{f^L}, \bar{f^H}) \gamma_\mu \left
  ( \begin{array}{cc}  
        \cos^2{\phi_f} T_3 - Q \sin^2{\theta_W} &
        \cos{\phi_f}\sin{\phi_f}T_3 \\  \cos{\phi_f}\sin{\phi_f}T_3 
        & \sin^2\phi_f T_3 - Q \sin^2\theta_W 
        \end{array} \right) \left( \begin{array}{c} f^L \\
      f^H \end{array}  \right) Z^\mu
\label{zcoupl}
\ee
where $T_3$ and $Q$ are the weak and electromagnetic charges of the ordinary
fermion.  The right-handed states, being weak singlets, couple to the $Z$
exactly as Standard Model right-handed fermions would.

The scalar boson $\Phi$ couples to the mass-eigenstate fermions according to
the Lagrangian term
\be
{\lambda_f}\, (\bar{f^L}, \bar{f^H})_{left}\left( \begin{array}{cc}
        -\cos{\phi_f}\,\sin{\phi_{f, right}} &
        \cos{\phi_f}\,\cos{\phi_{f, right}} \\  -\sin{\phi_f}\,\sin{\phi_{f,
            right}}   
        & \sin{\phi_f}\,\cos{\phi_{f, right}}
        \end{array} \right) \left( \begin{array}{c} f^L \\
        f^H \end{array}  \right)_{right} \Phi \ \ \ +\ h.c. \label{hcoupl}
\ee
where $\phi_{f, right}$ is the mixing angle for right-handed fermions.

A few notes about neutral-current physics are in order.  Flavor-conserving
neutral-current decays of the heavy states into light ones are possible (e.g.
$\mu^H \to \mu^L \nu_\mu \bar\nu_\mu$).  This affects the branching ratios in
heavy fermion decays and will be important in discussing searches for the
heavy states in Section 5.  Flavor-changing neutral (FCNC) processes are
absent at tree-level and highly-suppressed at higher order in the benchmark
model, due to the GIM mechanism mentioned earlier.  For example, we have
evaluated the fractional shift in the predicted value of $\Gamma(b \to s
  \gamma)$ by adapting the results in \cite{bbp}.  As we shall see in
  sections 3 and 4, electroweak data already constrain the mixings between
  ordinary and singlet fermions to be small and the masses of the heavy
  up-type fermion eigenstates to be large (so that the Wilson coefficients
  $c_7(m_f)$ that enter the calculation of $\Gamma(b\to s\gamma)$ are all in
  the high-mass asymptotic regime).  The shift in $\Gamma(b \to s \gamma)$ is
    therefore at most a few percent, which is well within the 10\% - 30\%
    uncertainty of the Standard Model theoretical predictions \cite{thepre}
    and experimental observations \cite{exobs}.

\section{General limits on mixing angles}
\label{sec:mixing}
\setcounter{equation}{0}

Precision electroweak measurements constrain the degree to which the
observed fermions can contain an admixture of weak-singlet exotic
fermions.  The mixing alters the couplings of the light fermions to
the $W$ and $Z$ from their Standard Model values, as discussed above,
and the shift in couplings alters the predicted values of many
observables.  Using the general approach of reference \cite{BGKLM}, we
have calculated how inclusion of mixing affects the electroweak
observables listed in Table 1.  The resulting expressions for these
leading (tree-level) alterations are given in the Appendix as
functions of the mixing angles.  We then performed a global fit to the
electroweak precision data to constrain the mixing angles between
singlet and ordinary fermions.  The experimental values of the
observables used in the fit and their predicted values in the Standard
Model are listed in Table 1.

To begin, we considered the benchmark scenario (called Case A,
hereafter) in which all electrically charged fermions have
weak-singlet partners \cite{origpaper}.  All of the electroweak
observables given in Table 1 receive corrections from fermion mixings
in this case.  We performed a global fit for the values of the 8
mixing angles of the fermions light enough to be produced at the
Z-pole: the 3 leptons, 3 down-type quarks and 2 up-type quarks.  At
95\% (90\%) confidence level, we obtain the following upper bounds on
the mixing angles:
\bea
\slp_e \leq  0.0024 \,\, (0.0020)\, ,\ \ \ \ \slp_\mu 
\!\!&\leq&\!\! 0.0030\,\, (0.0026)\, ,\ \ \slp\tau 
\leq 0.0030\,\, (0.0025) \nonumber \\
\slp_d \leq  0.015\,\, (0.013)\, , \ \ \ \ \ \ \ \slp_s 
\!\!&\leq&\!\! 0.015 \,\, (0.011)\, ,\ \ \ \ \ \slp_b \leq
0.0025\,\, (0.0019)\nonumber\\
\slp_u \leq  0.013\,\, (0.011)\, , \ \ \ \ \ \ \ \slp_c
\!\!&\leq&\!\! 0.020 \,\, (0.017)\, .\
\label{casea}
\eea
The 90\% c.l. limits are included to allow comparison with the slightly
weaker limits resulting from the similar analysis of earlier data in
reference \cite{nardi}.
  
The limits on the mixing angles are correlated to some degree.  For
 example, most observables that are sensitive to $d$ or $s$ quark
 mixing depend on $\slp_d + \slp_s$.  Indeed, repeating the global fit
 using the linear combinations $(\slp_d \pm \slp_s)/2$ yields a
 slightly stronger limit for the sum $(\slp_d + \slp_s)/2 \leq .0094$
 and a slightly looser one for the difference $-0.0071< (\slp_d -
 \slp_s)/2 \leq .0195$.  This turns out not to affect our use of the
 mixing angles to set mass limits in the next section of the paper:
 limits on the $d^H$ and $s^H$ masses arise from the more
 tightly-constrained $b$-quark mixing factor $\slp_b$ instead.

\begin{table}[bt]
\begin{center}
\begin{tabular}{|c|l|l|l|}\hline\hline
Quantity & Experiment & SM & Reference \\
\hline \hline
$\Gamma_Z$ & 2.4939 $\pm$ 0.0024 & 2.4958 & \protect\cite{LEPEWWG}\\
$\sigma_h$ & 41.491 $\pm$ 0.058 & 41.473 & \protect\cite{LEPEWWG}\\
$A_{\tau}(P_\tau)$ & 0.1431 $\pm$ 0.0045 & 0.1467 &
\protect\cite{LEPEWWG}\\
$A_{e}(P_\tau)$ & 0.1479 $\pm$ 0.0051 & 0.1467 & \protect\cite{LEPEWWG} \\
$A_{LR}$ & 0.1550 $\pm$ 0.0034 & 0.1467 & \protect\cite{PDG} (ref. 59) \\
$R_b$ & 0.21656 $\pm$ 0.00074 & 0.21590 & \protect\cite{LEPEWWG}\\
$R_c$ & 0.1735 $\pm$ 0.0044 & 0.1722 & \protect\cite{LEPEWWG}\\
$A_{FB}^b$ & 0.0990 $\pm$ 0.0021 & 0.1028 & \protect\cite{LEPEWWG}\\
$A_{FB}^c$ & 0.0709 $\pm$ 0.0044 & 0.0734 & \protect\cite{LEPEWWG}\\
${\cal{A}}_b$ & 0.867 $\pm$ 0.035 & 0.935 & \protect\cite{LEPEWWG}\\
${\cal{A}}_c$ & 0.647 $\pm$ 0.040 & 0.668 & \protect\cite{LEPEWWG}\\
$Q_W(Cs)$ & -72.41 $\pm$.25 $\pm$ .80 & -73.12 $\pm$ .06 &
\protect\cite{PDG} 
(refs. 48, 49)\\
$Q_W(Tl)$ & -114.8 $\pm$ 1.2 $\pm$ 3.4 & -116.7 $\pm$ .1 &
\protect\cite{PDG} 
(refs. 48, 49)\\
$R_e$ & 20.783 $\pm$ 0.052 & 20.748 & \protect\cite{LEPEWWG}\\
$R_\mu$ & 20.789 $\pm$ 0.034 & 20.748 & \protect\cite{LEPEWWG}\\
$R_\tau$ & 20.764 $\pm$ 0.045 & 20.748 & \protect\cite{LEPEWWG}\\
$A_{FB}^e$ & 0.0153 $\pm$ 0.0025 & 0.01613 & \protect\cite{LEPEWWG}\\
$A_{FB}^\mu$ & 0.0164 $\pm$ 0.0013 & 0.01613 & \protect\cite{LEPEWWG}\\
$A_{FB}^\tau$ & 0.0183 $\pm$ 0.0017 & 0.01613 & \protect\cite{LEPEWWG}\\
$A_{FB}^s$ & 0.118 $\pm$ 0.018 & 0.1031 $\pm$.0009 & \protect\cite{PDG}\\
$M_W$ & 80.39 $\pm$ 0.06 & 80.38 & \protect\cite{Lweb}\\
$g_{eV}(\nu e \rightarrow \nu e)$ & -0.041$\pm$ 0.015& -0.0395$\pm$.0005 & 
\protect\cite{PDG}\\
$g_{eA}(\nu e \rightarrow \nu e)$ & -0.507 $\pm$ 0.014& -0.5064$\pm$.0002& 
\protect\cite{PDG}\\
$g_L^2(\nu N \rightarrow \nu X)$ & 0.3009$\pm$0.0028 & 0.3040 $\pm$ .0003& 
\protect\cite{PDG}\\
$g_R^2(\nu N \rightarrow \nu X)$ & 0.0328 $\pm$ 0.0030 & 0.0300 &
\protect\cite{PDG}\\
$R_{\pi}$&(1.230 $\pm$ .004)$\times 10^{-4}$ & (1.2352 $\pm$ .0005)$\times
10^{-4}$ & \protect\cite{PDG}, \protect\cite{Sirl}\\
$R_{\tau}$ & (1.347  $\pm$ .0082)$\times 10^{6}$ & 1.343 $\times 10^{6}$ &
\protect\cite{PDG}, \protect\cite{Pich} \\
$R_{\mu \tau}$ & (1.312  $\pm$ .0087)$\times 10^{6}$ & 1.304 $\times
10^{6}$ &
\protect\cite{PDG}, \protect\cite{Pich}\\
\hline\hline
\end{tabular}
\end{center}
\caption{Data used in fits to constrain mixing angles: experimentally 
measured electroweak observables and their values within the 
Standard Model.} 
\label{Pred}
\end{table}

We, similarly, placed limits on the relevant mixing angles for three
scenarios in which only third-generation fermions have weak-singlet partners.
In Case B where the top quark, bottom quark and tau lepton have partners, the
12 sensitive observables are $\Gamma_Z, \sigma_h, R_{b,\,c,\,e,\,\mu,\,\tau},
A_{FB}^{b,\tau}, {\cal{A}}_b,$ and $R_{e\tau, \mu\tau}$.  The resulting
 95\% (90\%) confidence level limits on the bottom and tau mixing angles are
\be
\slp_\tau \leq  0.0018\,\, (0.0014)\, ,\ \ \ \ \ \slp_b \leq 
0.0013\,\, (0.00084)
\label{caseb}
\ee
In Case C, where only the top and bottom quarks have partners, the nine
affected quantities are $\Gamma_Z, \sigma_h, R_{b,\,c,\,e,\,\mu,\,\tau},
A_{FB}^b,$ and ${\cal{A}}_b$.  The sole constraint is
\be
\slp_b \leq \, 0.0013\,\, (0.00084)
\label{casec}
\ee
In Case D, where only the tau leptons have partners, only the six quantities
$\Gamma_Z, \sigma_h,$ $R_{\tau}, A_{FB}^\tau,$ and $R_{e\tau, \mu\tau}$ are
sensitive, and the limit on the tau mixing angle is
\be
\slp_\tau \leq  0.0020\,\, (0.0016)
\label{cased}
\ee

These upper bounds on the mixing angles depend only on which fermions have
weak partners, and not on other model-specific details.  They apply broadly
to theories in which the low-energy spectrum is that of the Standard Model plus
weak-singlet fermions.

\section{From mixing angles to mass limits}
\label{sec:masses}
\setcounter{equation}{0}

The constraints on the mixing between
the ordinary and exotic fermions imply specific lower bounds on the masses of
the heavy fermion mass eigenstates (\ref{hemass}).  We will 
extract mass limits from mixing angle limits first in the general case
\cite{origpaper} in which all charged fermions have singlet partners, and then
in scenarios where only the third generation fermions do.

\subsection{Case A: all generations mix with singlets} 

Because the heavy fermion masses $m_f^H$ depend on $v\lambda_f$, $M_f$, and
$\sm_f$, we must determine the allowed values of all three of these
quantities in order to find lower bounds on the $m_f^H$.  For the three
fermions of a given type, (e.g. $e$, $\mu$, $\tau$), the values of
$\lambda_f$ and $M_f$ are common.  The different values of $m^L_f$ arise from
differences among the $\sm_f$, and the form of equation (\ref{limass}) makes it
clear that larger values of $\sm_f$ correspond to larger values of $m^L_f$.

How can we ensure that the third-generation fermion in the set gets a large
enough mass ?  If we set $\sm_f$ to the largest possible value, $\sm_f =
M_f$, there is a minimum value of $v \lambda_f$ required to make
$m^L_f$ large enough.  A smaller value of $\sm_f$ would require a still
larger value of $v \lambda_f$ to arrive at the same $m^L_f$.  In other words,
starting from (\ref{limass}), and recalling $v \lambda_f < M_f$ we find
\be
v \lambda_f \geq \sqrt{2} m^L_{f3}
\label{vlamlim}
\ee
where ``$f3$'' denotes the third-generation fermion of the same type as
``$f$'' (e.g. if ``$f$'' is the electron, ``$f3$'' is the tau lepton).  The
specific limits for the three types of charged fermions are:
\be 
v \lambda_\ell \geq 2.5\ {\rm GeV}, \ \ \ \ v \lambda_D \geq 6.0\ 
{\rm GeV}\ \ \ \  v \lambda_U \geq 247\ {\rm GeV} 
\label{vlamlimsp}
\ee

Knowing this allows us to obtain a rough lower bound on the heavy fermion
mass eigenstates.  Since we require $M_f \geq v \lambda_f$ and since the
smallest possible value of $\sm_f$ is zero, we can immediately apply
(\ref{vlamlim}) to (\ref{hemass}) and find
\be
(m_f^H)^2 \gae 4 (m_{f3}^2) - (m_f^L)^2
\label{weaklim}
\ee
For instance, the mass of the heavy top eigenstate must be at least 
\be
m_t^H \gae \sqrt{3} m_t^L \approx 300 {\rm GeV}
\ee

We can improve on these lower bounds in the following way.  Because
$(m_f^H)^2$ is a monotonically increasing function of $(v \lambda)^2$,
the minimum $v \lambda_f$, found above, yields the lowest possible value of
$m^H_f$.  Thus, if we know what value of $\sm_f$ should be used
self-consistently with the smallest $v\lambda_f$, we can use (\ref{hemass})
to obtain a more stringent lower bound on $m_f^H$.  The appropriate values
\bea
\sm_f & = & M_f \ \ \ \ \ {\rm (3rd\ generation)} \label{trigen} \\
\sm_f & = & m^L_f 
       \sqrt{ M_f^2 + v^2 \lambda_f^2 \over{v^2 \lambda_f^2 -
           (m^L_f)^2}}\ \ \ \ \  {\rm (1st\ or\ 2nd\ generation)} 
\label{liigen}
\eea
follow from our previous discussion and from inverting equation  (\ref{limass}),
respectively.  Because $\sm_f << M_f$ for the first and second generation
fermions, our previous lower bound on $m_f^H$ for those generations is not
appreciably altered.  For the third generation we obtain the more
restrictive
\be
(m_f^H)^2 \gae 5 (m_{f3}^L)^2 
\label{medlim}
\ee
so that, for example,
\be
m_t^H \gae \sqrt{5} m_t^L \approx 390\ {\rm GeV}\ \ .
\label{mtoprev}
\ee

We can do still better by invoking our precision bounds on the mixing angles
$\slps_f$.  Recalling $v
\lambda_f < M_f$ and $\sm_f \le M_f$, allows us to approximate our expression
(\ref{sineq}) for the mixing angle as
\be
\slps_f \approx 
{{v \lambda_f M_f}\over {M_f^2 + \sm_f^2 - v^2 \lambda_f^2}} \ .
\ee
Further simplification of this relation depends on the generation to which
fermion $f$ belongs.  For example, among the charged leptons, $\sm_e$ and
$\sm_\mu$ are far smaller than $M_\ell$, while $\sm_\tau$ could conceivably
be of the same order as $M_\ell$.  Thus the limits on the leptons' mixing
angles imply
\be
M_\ell \geq {\rm Max} \left[ {v \lambda_\ell \over {2 \slps_\tau}} \, , {v \lambda_\ell \over {\slps_\mu}}\, , {v \lambda_\ell \over {\slps_e}} \right]
\label{mallthree}
\ee
The strongest bound on $M_\ell$ comes from $\sin\phi_e$; that
for $M_D$, from $\sin\phi_b$; that for $M_U$, from $\sin\phi_u$: 
\be
M_\ell \geq {v \lambda_\ell \over {\slps_e}} \ge 51\, {\rm GeV},\ \ \ \ \ \ 
\ 
M_U \geq {v \lambda_U \over {\slps_u}} \ge 2.2\,{\rm TeV},\ \ \ \ \  \  
M_D \geq {v \lambda_D \over {2 \slps_b}} \ge 60\, {\rm GeV}
 \label{melllim} 
\ee

Combining those stricter lower limits on $M_f$ with our bounds
(\ref{vlamlim}) on $v\lambda_f$ and our expression for the heavy fermion mass 
(\ref{hemass}) gives us a lower bound on the $m^H_f$ for each fermion flavor.
For the third generation fermions we use (\ref{trigen}) for the value of
$\sm_f$ and obtain the 95\% c.l. lower bounds
\bea
m^H_\tau & \ge & m^L_\tau \sqrt{1 + 4/\slp_e}\ \ge\ 73 \ {\rm GeV} 
\label{mhtaua} \\
m^H_t & \ge &  m^L_t \sqrt{1 + 4/\slp_u}\ \ge\ 3.1 \ {\rm TeV} \label{mhtopa}\\
m^H_b & \ge &  m^L_b \sqrt{1 + 1/\slp_b}\ \ge\ 86 \ {\rm GeV}\ \ .
\eea
For the lighter fermions, we use equation (\ref{liigen})
for the $\sm_f$.  Since $\sm_f << M_f$ in these cases, we find 
\bea
m^H_e, m^H_\mu & \gae &  m^L_\tau \sqrt{2 + 2/\slp_e}
               \ \ \ge\ \ 51\, {\rm {GeV} \label{mhem}}\\ 
m^H_u, m^H_c & \gae &  m^L_t \sqrt{2 + 2/\slp_u} 
               \ \ \ge\ \ \ 2.2\, {\rm {TeV}} \label{mhuc}\\
m^H_d, m^H_s & \gae &  m^L_b \sqrt{2 + 1/2\slp_b} 
               \ \ \ge\ \ \ 61\, {\rm {GeV}}\ \ .\label{mhds}
\eea

\begin{figure}
\rotatebox{90}{\resizebox{8cm}{14cm}{\includegraphics{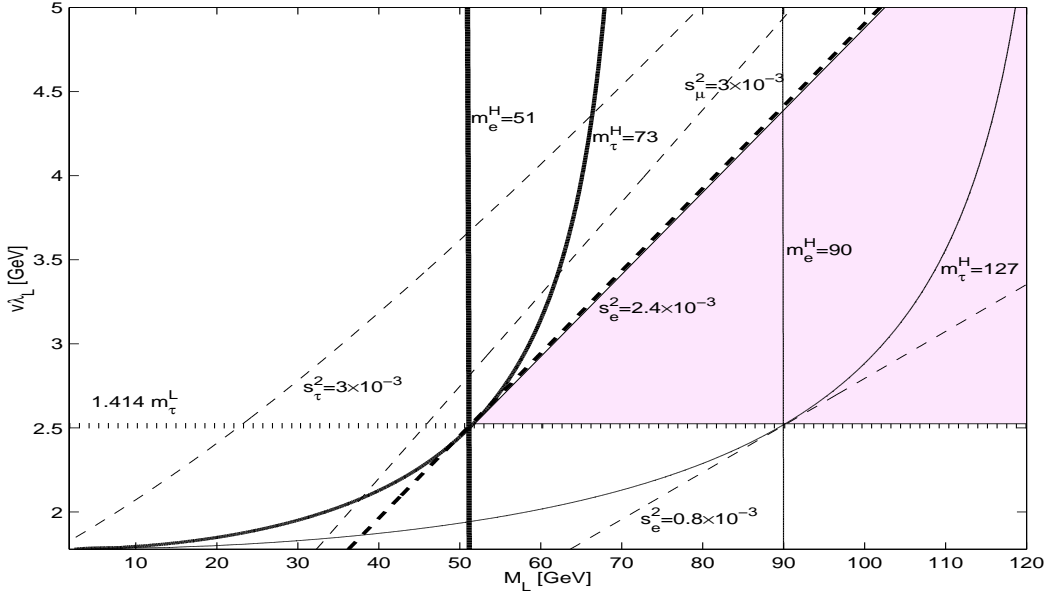}}}
\caption[lepton]{\small A graphical representation of the coupling and
  mass limits for the $e^H$ and $\tau^H$ states in the $M_\ell$ vs.
  $v\lambda_\ell$ plane.  The shaded region indicates the experimentally
  allowed region of the parameter space.   The quantities on the axes are
  flavor-blind; separate curves for different lepton flavors are shown.  Each
  dashed curve shows the upper bound on $\slp_\ell$ for one lepton and
  excludes the region to its left.  Each solid curve shows the lower bound on
  $m^H_\ell$ for one lepton (the limits for $e$ and $\mu$ are identical, as
  discussed in the text and equation(\protect\ref{mhem}). The horizontal
  dotted line indicates the minimum allowed value of $v\lambda_\ell$
  (\protect\ref{vlamlim}).}
\label{leptonfig}
\end{figure}

\begin{figure}
\rotatebox{90}{\resizebox{8cm}{14cm}{\includegraphics{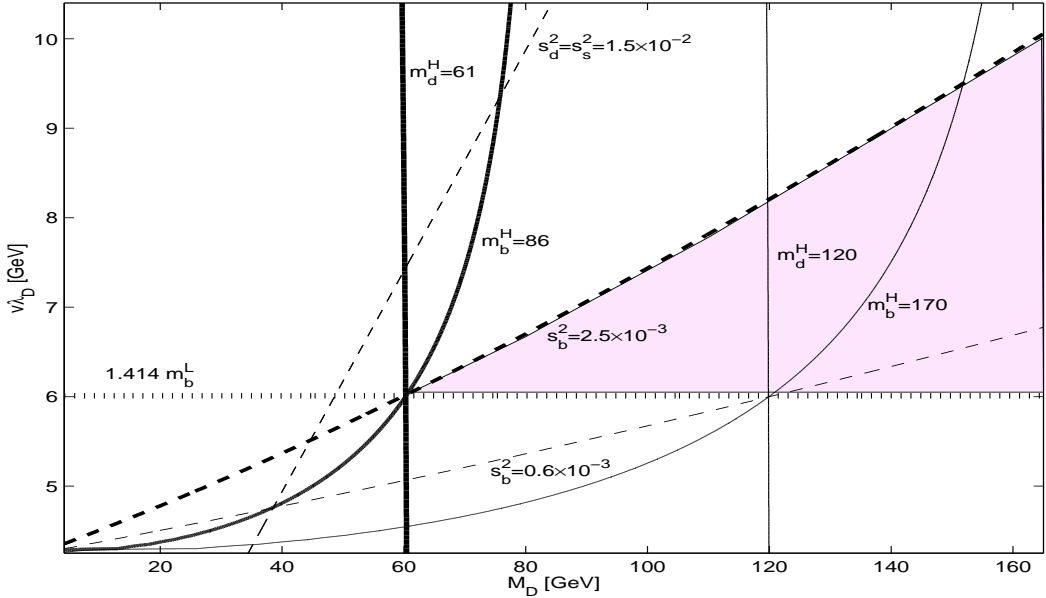}}}
\caption[down]{\small A graphical representation of the mass limits for
  the $d^H$ and $b^H$ states in the $M_D$ vs. $v\Lambda_D$ plane.  The shaded
  region indicates the experimentally allowed region of the parameter space
  The quantities along the axes are flavor-blind; separate curves for
  different quark flavors are shown.  Each dashed curve shows the upper bound
  on $\slp_D$ for one quark and excludes the region to its left.  Each solid
  curve shows the lower bound on $m^H_D$ for one quark (the limits for $d$
  and $s$ are identical, as discussed in the text and equation
  (\protect\ref{mhds}). The horizontal dotted line indicates the minimum
  allowed value of $v\lambda_D$ (\protect\ref{vlamlim}).}
\label{downfig}
\end{figure}

The mass limits for the heavy leptons and down-type quarks are also
represented graphically in figures \ref{leptonfig} and \ref{downfig}.
In figure \ref{leptonfig}, which deals with the leptons, the axes are
the flavor-universal quantities $M_\ell$ and $v \lambda_\ell$.  The
shaded region indicates the experimentally allowed region of the
parameter space. The lower edge of the allowed region is delimited by
the lower bound on $v \lambda_\ell$ of equation (\ref{vlamlimsp}), as
represented by the horizontal dotted line.  The left-hand edge of the
allowed region is demarked by the upper bound on the electron mixing
factor, $\slp_e$, as shown by the dashed curve with that label.  The
form of this curve, $\slp_e(M_\ell, v\lambda_\ell) = 0.0024$, was
obtained numerically by using equation (\ref{lameq}) for $m_e^L$ to
put the unknown $\sm_e$ in terms of $M_\ell$, $v \lambda_\ell$ and the
observed mass of the electron ($m_e^L = .511$ MeV) and inserting the
result into equation (\ref{sineq}).  The curves for the muon and tau
mixing angles were obtained similarly, but provide weaker limits on
the parameter space (as shown by the dashed curves labeled $\slp_\mu$,
and $\slp_\tau$).  The lowest allowed values of the heavy fermion
masses $m_{e,\mu}^H$ and $m^H_\tau$ are those whose curves intersect
the tip of the allowed region; these are shown by the solid curves,
obtained numerically by using equation (\ref{lameq}) to replace the
unknown $\sm_e, \sm_\tau$ by the known $m^L_e, m^L_\tau$ in our
expressions for $m^H_{e}$ and $m^H_\tau$.  Figure \ref{downfig} shows
the analogous limits on the mixing angles and heavy-eigenstate masses
for the down-type quarks.

We can also construct a plot of the allowed region of $M_U$ vs. $v
\lambda_U$ parameter space.  The lower edge comes from the lower bound
on $v\lambda_U$ and the left-hand edge, from the upper bound on $\slps_u$.
We can then use the known value of $m^L_t$ to calculate the size of the
 top quark mixing factor $\slp_t$ at any given point
in the allowed region.  Numerical evaluation reveals
\be
\slp_t \leq  0.013 \,\, (0.011)\label{consist}  
\ee
at 95\% (90\%) c.l. This is a limit on top quark mixing imposed by
self-consistency of the model.

In section \ref{sec:dir-prod}, we will compare the mass limits just extracted
from precision data with those derived from searches for direct production of
new fermions at the LEP II and Tevatron colliders.  The lower bounds on the
masses of the heavy down-type quarks or charged leptons admit the possibility
of those particles' being produced at current experiments.  The heavy up-type
quarks are too massive to be even singly produced at existing
colliders.

\subsection{Cases B, C, and D: third-generation fermions mix with singlets}

If only third-generation fermions have weak-singlet partners, there are a few
differences in the analysis that yields lower bound on heavy eigenstate
masses.  All follow from the fact that the lower bounds on the $M_f$ (as in
equation (\ref{mallthree})) can no longer come from precision limits on the
mixing angles of 1st or 2nd generation fermions (since those fermions no
longer mix with weak singlets).

To obtain the precision bounds on the masses of $b^H$ and $\tau^H$, we start
by writing the lower limits on $M_\ell$ and $M_D$ that come from the
mixing angles:
\be
M_\ell \geq {v \lambda_\ell \over {2 \slps_\tau}} \, ,\ \ \ \ \ \ 
M_D \geq {v \lambda_Dl \over {2 \slps_b}}\ \ .
\ee
The factor of 2 in the denominator arises because the mixing angles belong to
a third-generation fermion (so that $\sm_f = M_f$). We therefore find 
\bea
m^H_\tau & \ge & m^L_\tau \sqrt{1 + 1/\slp_\tau}\label{mhtaubcd}\\
m^H_b & \ge &  m^L_b \sqrt{1 + 1/\slp_b}.\label{mhbbcd}
\eea
  In Case B where all third-generation fermions mix with weak-singlet
fermions, the mixing angle limits (\ref{caseb}) based on the twelve
sensitive observables yield 95\% c.l.  lower bounds
\bea
m^H_\tau &\ge& 42\ {\rm GeV}\label{mhtaub}\\
m^H_b & \ge& 119\ {\rm GeV}
\eea
In Case C, where only third-generation quarks have partners, (\ref{casec}) which was obtained by a fit to the nine aaffected observables,
gives 
\be
m^H_b \ge 119\ {\rm GeV}
\label{limcb}
\ee
while in Case D, where only tau leptons have partners, (\ref{cased})
based on six affected precision electroweak quantities implies
\be
m^H_\tau \ge 40\ {\rm GeV}\label{mhtaud}
\ee
Compared with the limits in Case A, we see that the lower bound on $m^H_b$ is
strengthened because the precision limit (\ref{caseb},\,\ref{casec}) on
$\slp_b$ is more stringent.  In contrast, the lower bound on $m^H_\tau$ is
weakened because the bound now depends on a third-generation instead of a
first-generation mixing angle: equation (\ref{mhtaubcd}) is approximately
$m_\tau^H \geq m^L_\tau / \slps_\tau$ whereas equation (\ref{mhtaua}) was
roughly $m_\tau^H \geq 2 m^L_\tau / \slps_e$.

Note that in theories where the top is the only
up-type quark to have a weak-singlet partner, such as Cases B and C, the only
bound on $m^H_t$ comes from equation (\ref{mtoprev}).  While this is far weaker
than the limit in Case A, it still ensures that the heavy top eigenstate is
too massive to have been seen in existing collider experiments, even if singly
produced.  

\section{Limits on direct production of singlet fermions}
\label{sec:dir-prod}
\setcounter{equation}{0}

While interpreting the general mixing angle limits in terms of mass limits
requires specifying an underlying model structure, it is also possible to set
more general mass limits by considering searches for direct production of the
new fermions.  The LEP experiments have published limits on new sequential
charged leptons \cite{OPAL-lepton}\cite{DELPHI-lepton}; the Tevatron
experiments have done the same for new quarks
\cite{D0-quark}\cite{CDF-quark}.  In this section, we adapt the limits to
apply to scenarios in which the new fermions are weak singlets rather than
sequential.

\subsection{Decay rates of heavy fermions}

\begin{figure}
\begin{center}
{\includegraphics{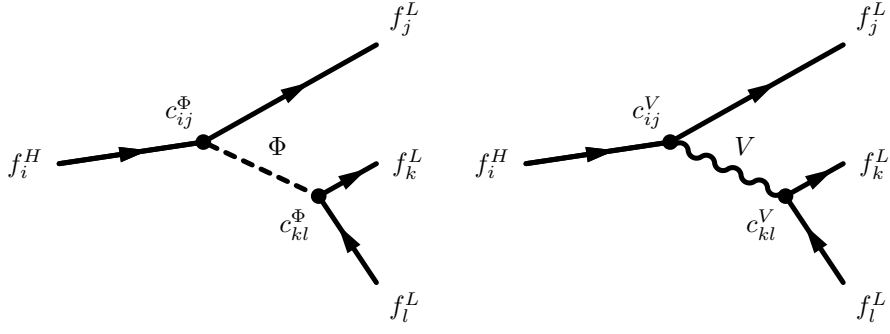}}
\end{center}
\caption[down]{\small Scalar ($\Phi$) and 
  weak boson (V $\equiv$ Z or W) decay modes of a heavy fermion ($f^H$).} 
\label{testpaperfig}
\end{figure}

A heavy fermion decays preferentially to a light fermion\footnote{Even where
  a heavy fermion is kinematically allowed to decay to another heavy fermion,
  the rate is doubly-suppressed by small mixing factors $(\slps_f)$ and,
  consequentially, negligible.} plus a Z, W, or $\Phi$ boson which
subsequently decays to a fermion-antifermion pair (see figure
\ref{testpaperfig}) \footnote{In this section we confine our analysis to
  relatively light scalars, with mass below 130 GeV.  For heavier scalar one
  should include the scalar decays to W and Z pairs \cite{hscaldcy} and the
  resulting 5-fermion final states of heavy fermion decays.  We expect this
  to yield only a small change in the results of our quark-sector analysis
  and essentially no alteration in our results for heavy lepton decays, due
  to the large kinematic suppression when $m_\Phi >> m^H_\ell \sim M_W$.}
  
At tree-level, and neglecting final state light fermion masses, we obtain
 the following partial rates for vector
boson decay modes of the heavy fermions
\bea
\!\Gamma(f^H_i\rightarrow f^L_j V)&=&\sum_{k,l}\Gamma(f^H_i\rightarrow
f^L_j V\rightarrow f^L_jf^L_kf^L_l) \label{decayyZ/W}\\
&=&\sum_{k,l}{(c^V_{ij}\,c^V_{kl})^2
  \over {3 \, \pi^3 \,  2^8}}\, m_{f_i}^H \, F \bigg{[}{\bigg{(}{m_{f_i}^H 
\over M_V}\bigg{)}}^2\, , \, {\Gamma_V
  \over M_V}{\bigg {]}}\nonumber 
\eea
where V represents a Z or W boson, while $\Gamma_V$ and $M_V$ are,
respectively, the vector boson's decay rate and mass.  Function $F(x,y)$ is
presented in appendix \ref{sec:appx2}.  The vertex factors
$c^V_{ij}$ ($c^V_{kl}$) are, as shown in figure \ref{testpaperfig}, the
$f^H_i f^L_j V$ ($f^L_k f^L_l V$) couplings which may be read from 
equations (\ref{lwco}) -- (\ref{zcoupl}).

Our results for the charged-current decay mode agree with those presented in
integral form in \cite{Wmode}.  Moreover, equation yields the standard
asymptotic behaviors in the limit of heavy fermion masses far above or far
below the electroweak bosons' masses (see appendix \ref{sec:appx2}).  Since
some of our heavy fermions can, instead, have masses of order 80-90 GeV, we
use the full result (\ref{decayyZ/W}) in our evaluation of branching fractions
and search potentials.

Similarly, we find the partial rate for the scalar decay mode to be
\bea
\!\Gamma(f^H_i\rightarrow f^L_j\, \Phi)&=&\sum_{k,l}\Gamma(f^H_i\rightarrow
f^L_j \,
\Phi\rightarrow f^L_jf^L_kf^L_l) \label{decayH} \\
&=&\sum_{k,l}{(c^\Phi_{ij}\,c^\Phi_{kl})^2
  \over {\pi^3 \,  2^{10}}}\, m_{f_i}^H \, G \bigg{[}{\bigg{(}{m_{f_i}^H 
\over M_\Phi}\bigg{)}}^2\, , \, {\Gamma_\Phi
\over M_\Phi}{\bigg {]}}\nonumber 
\eea
where $\Gamma_\Phi$ and $M_\Phi$ are the decay rate and the mass of the scalar
boson, $\Phi$.  Function $G(x,y)$ and additional
details are given in appendix \ref{sec:appx2}.  The vertex factors $c^\Phi_{ij}$
($c^\Phi_{kl}$) are, as indicated in figure \ref{testpaperfig},  the
$f^H_i f^L_j \Phi$ ($f^L_k f^L_l \Phi$) couplings which may be read off of
equation (\ref{hcoupl}).

We have numerically evaluated the couplings of the light fermions to the
scalar\footnote{To evaluate the mixing among right-handed fermions which
  appears in the fermion-scalar couplings, we derive a relation analogous to
  (\ref{analggy}) and apply equation $\ref{limass}$ so that
  $\srp$ is written in terms of known light fermion masses, the $M_f$ and the
  $v\lambda_f$.}, Z, and W as 
functions of the $M_f$ and the $v\lambda_f$.  In the region of the model
parameter space that is allowed by precision electroweak measurements, we
find that these couplings are within 1\% of their Standard Model values.
Therefore, in this section of the paper, we approximate the $\Phi f f$ and $V
f f$ couplings for the light fermions by the Standard Model values.  This
allows us to express our results for branching fractions and searches in the
simple $M_f$ vs. $v\lambda_f$ planes for the up, down, and charged-lepton
sectors.  In this approximation, the recent LEP lower bound on the mass of
the Higgs boson \cite{L3limit}, $M_H\geq95.3$ GeV, applies directly to the
mass of the $\Phi$ scalar in our model:
\be
M_\Phi \geq 95.3\ {\rm GeV}
\label{lephiggs}
\ee

\begin{figure}
\rotatebox{90}{\resizebox{8cm}{14cm}{\includegraphics{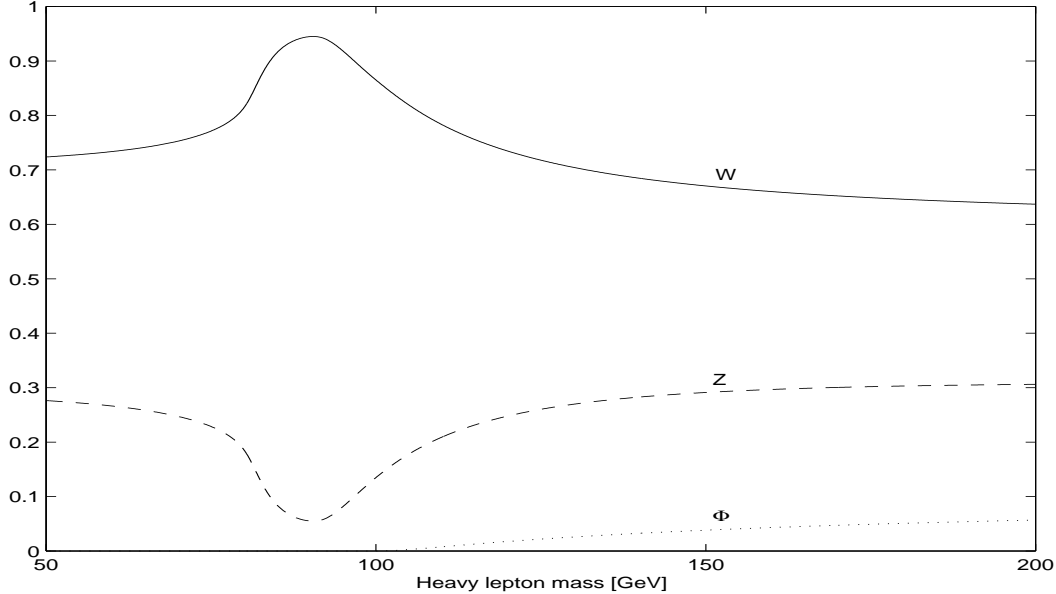}}}
\caption[lbranch]{\small Branching ratios in the heavy lepton
  sector: $B(\ell^H \to \ell^L X)$ where $X$ is W, Z, or $\Phi$. We set
  $v\lambda_\ell=2\,m_\tau^L$ and $M_\Phi=100$ GeV.}
\label{lbranchfig}
\end{figure}

The branching ratios for the decays of the heavy leptons are effectively
flavor-universal, i.e. the same for $e^H$, $\mu^H$, and $\tau^H$.  The
charged-current decay mode dominates; decays by $Z$ emission are roughly half
as frequent and decays by $\Phi$ emission contribute negligibly for $m^H_f\leq
M_\Phi$.  In the limit where the heavy lepton masses $m^H_\ell$ are much
larger than any boson mass, the branching ratios for decays to W, Z, and $\Phi$
approach $60.5\%$, $30.5\%$, and $9\%$, respectively.  The branching
fractions for heavy lepton decays are shown in figure
\ref{lbranchfig} as a function of heavy lepton mass $m^H_\ell$, with $M_\Phi$
fixed at 100 GeV and $v\lambda_\ell$ set equal to $2 m_3^L$.
As the branching ratios have little dependence on the small mixing factors
$\slps_f$ (as we argue in more detail in the following subsection), they are
also insensitive to the value of $v\lambda_f$.

The branching fractions for decays of the heavy down-type quarks display a
significant flavor-dependence.  Those for the $d^H$ and $s^H$ are essentially
identical and resemble the branching fractions for the heavy leptons.
However, charged-current decays of $b^H$ with a mass less than 255 GeV (the
threshold for decay to an on-shell top and W) are doubly Cabbibo-suppressed,
so that the $b^H$ branching ratios do not resemble those of the other
down-type quarks.  Generally speaking, a $b^H$ of relatively low mass decays
almost exclusively by the process $b^H \to Z b^L$.  For $m^b_H$ larger than
255 GeV, the decay $b^H \to t^L W$ dominates and the Z-mode branching
fraction is only about half as large.  If $m_b^H$ is above $M_\Phi + m_b^L$ but
below 255 GeV, the scalar decay mode becomes significant (in agreement with
reference \cite{Hmode}).  If the scalar mass lies above 255 GeV, the scalar
decay mode is much less important.  In the asymptotic regime, where $m_D^H$
is much greater than $m_t$ or any boson's mass, the branching ratios for
decays to W, Z, and $\Phi$ approach $49\%$, $25\%$, and $26\%$, respectively.

\begin{figure}
\rotatebox{90}{\resizebox{8cm}{14cm}{\includegraphics{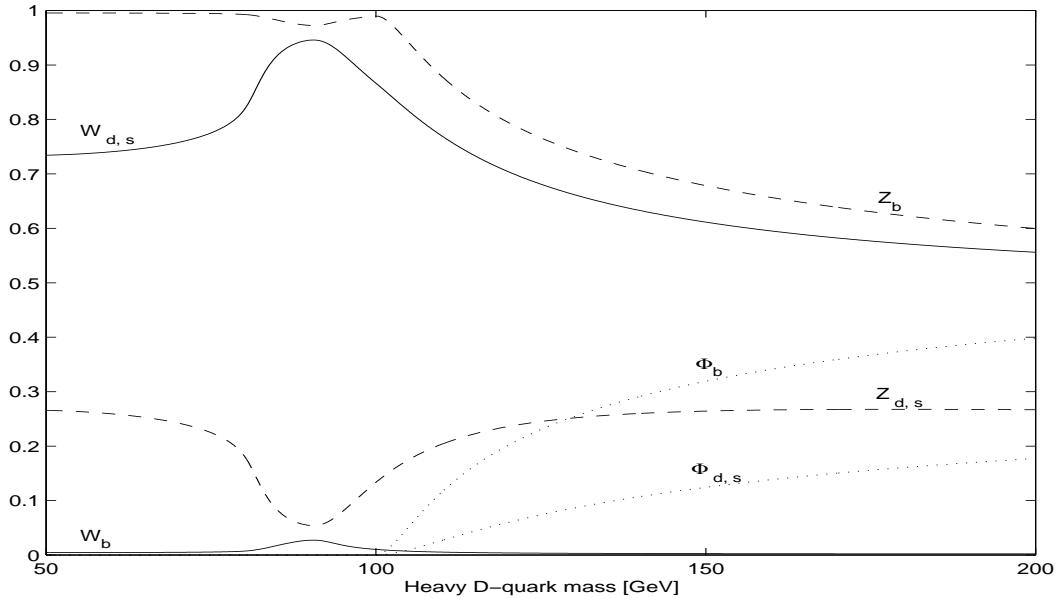}}}
\caption[bbranch]{\small Branching ratios in the heavy down-quark
  sector: $B(D^H \to D^L X)$ where $X$ is W, Z, or $\Phi$.  Subscripts d, s,
  and b denote the heavy quarks' flavor. We set $v\lambda_D=2\,m_b^L$ and
  $M_\Phi=100$ GeV.}
\label{bbranchfig}
\end{figure}

\subsection{Heavy leptons at LEP II}

The LEP II experiments have searched for evidence of new sequential leptons,
working under the assumptions that the new neutral lepton $N$ is heavier than
its charged partner $L$ and that $L$ decays only via charged-current mixing
with a Standard Model lepton (i.e. $B(L \to \nu_\ell W^*) =$ 100\%).  Recent
limits from the OPAL experiment at $\sqrt{s} = 172$ GeV
\cite{OPAL-lepton} and from the DELPHI experiment at $\sqrt{s} = 183$ GeV
\cite{DELPHI-lepton} each set a lower bound of order 80 GeV on the mass of
a sequential charged lepton.

To illustrate how the LEP limits may be applied to our weak-singlet
fermions, we review OPAL's analysis.  The OPAL experiment searched for
pair-produced charged sequential leptons undergoing
charged-current decay:
\be
e^+ e^- \to L^+ L^- \to \nu_\ell \bar{\nu_\ell} W^+ W^-
\label{opalpath}
\ee
Their cuts selected final states in which at least one of the $W^*$ bosons
decayed hadronically.  Events with no isolated lepton were required to have
at least 4 jets and substantial missing transverse momentum; those with one
or more isolated leptons were required to have at least 3 jets, less than 100
GeV of visible energy, and substantial missing transverse momentum.  The
efficiencies for selecting signal events were estimated at 20-25\% by Monte
Carlo.  With 1 candidate event in the data set and the expectation of 3
Standard Model background events, OPAL excluded, at 95\% c.l., sequential
leptons of mass less than 80.2 GeV, as these would have contributed least 3
signal events to the data.

\begin{figure}
\rotatebox{90}{\resizebox{8cm}{14cm}{\includegraphics{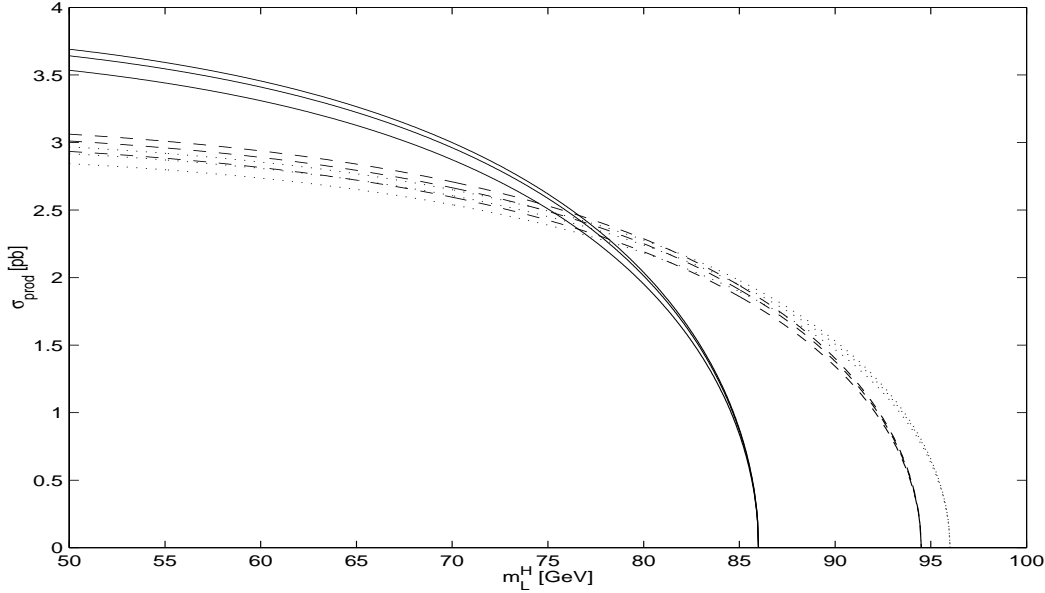}}}
\caption[rts]{\small Production cross-section for a heavy lepton that is
  mostly weak-singlet as a function of lepton mass and mixing angles.
Each
  family of curves represents one value of $\sqrt{s}$ (solid = 172 GeV,
  dashed = 189 GeV, dotted = 192 GeV).  The separate curves within each
  family show the effect of changing the value of the small mixing angles
  (top curve: $\slp_\ell$ = 0.1 for all leptons, middle curve: $\slp_\ell
=
  0$ for all leptons, bottom curve: $\slp_e$ = 0, $\slp_{\mu,\tau}$ =
0.1).}
\label{rtsfig}
\end{figure}

The heavy leptons in the models we are studying have different weak quantum
numbers than those OPAL sought.  This alters both the production rate and the
decay paths of the leptons.  The production rate of the $\ell^H$ should be
larger than that for the sequential leptons. The pure QED contribution is the
same, as the heavy leptons have standard electric charges; the
weak-electromagnetic interference term is enhanced since the coupling to the
$Z$ is roughly $\sin^2\theta_W > 0$ rather than $\sin^2\theta_W - 0.5 < 0$ as
in the Standard Model.  By adapting the results of reference
\cite{production} to include the couplings appropriate to our model, we have
evaluated the production cross-section for heavy leptons at LEP II.  Our
results are shown in figure \ref{rtsfig} as a function of heavy lepton mass
for several values of $\sqrt{s}$ and lepton mixing angle.

On the other hand, the likelihood that our heavy leptons decay to final
states visible to OPAL is less than it would be for heavy sequential leptons.
In events where both of the produced $\ell^H$ decay via charged-currents,
about 90\% of the subsequent (standard) decays of the $W$ bosons lead to the
final states OPAL sought -- just as would be true for sequential leptons.
But the heavy leptons in our model are not limited to charged-current decays.
In events where one or both of the
produced $\ell^H$ decay through neutral currents, the result need not be a
final state visible to OPAL.  If there is one W and one Z in the intermediate
state, about 36\% of the events should yield final states with sufficient
jets, isolated leptons and missing energy to pass the OPAL cuts.  At the
other extreme, if both $\ell^H$ decay by $\Phi$ emission, there will be
virtually no final states with sufficient missing energy, since $\Phi$ decays
mostly to $b\bar{b}$.  The other decay patterns lie in between; for
intermediate $ZZ$ ($\Phi Z$, $\Phi W$) we expect 28\% (19\%, 30\%) of the
events to be visible to OPAL.  The total fraction of pair-produced heavy
leptons that yield appropriate final states is the sum of these various
possibilities:
\bea
B_{decay} &=& 0.896\,(B_W)^2 + 0.280\,(B_Z)^2 + 2\,(\,0.361\, B_W \cdot
B_Z \nonumber \\ &+& 0.190\,B_Z \cdot B_\Phi \, + \, 0.306 \, B_\Phi \cdot B_W
\,)
\label{Bdecay}
\eea
where $B_{W}$, $B_{Z}$, and $B_{\Phi}$ are the heavy lepton branching fractions
for the W, Z and scalar decay modes respectively, as calculated in section 5.1
(and shown in Figure 4).

\begin{figure}
\rotatebox{90}{\resizebox{8cm}{14cm}{\includegraphics{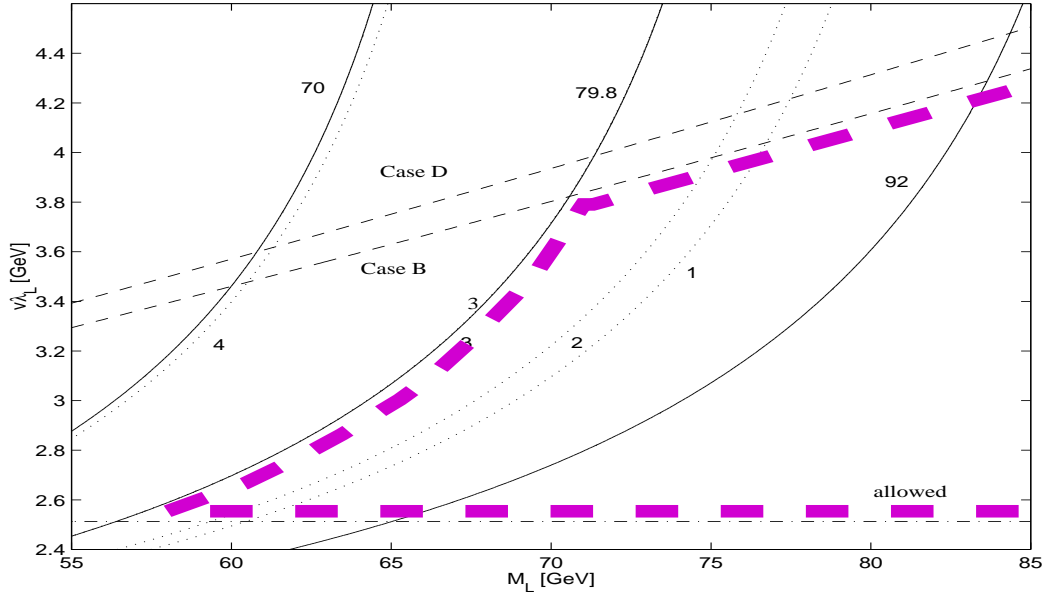}}}
\caption[OPALlim]{\small Lower bounds on heavy lepton mass in models with one
  flavor of heavy lepton (Cases B and D).  The heavy dashed curve encloses
  the region of the plot allowed by the combination of electroweak and direct
  search data.  The limits on $\sin{\phi_{\tau}}^2$ (dashed lines) and
  $v\lambda_{\ell}=m^L_{\tau}\sqrt{2}$ (dot-dash line) are as in Figure 1.
  Solid lines are contours of constant heavy lepton mass $m^H_\tau$; dotted
  curves are contours of constant numbers of OPAL signal events.  The lower bound
  the lepton mass comes from the overlap of the limiting $N=3$ dotted line with the
  $m^H_\tau=79.8 \ {\rm GeV}$ solid line.   In calculating branching fractions, the scalar mass was set
  to 100 GeV.}
\label{OPALlimfig}
\end{figure}

In models (cases B and D) where there is only one species of heavy lepton
($\tau^H$), setting a mass limit is straightforward.  We note that
\be
\sigma_{production}\cdot B_{decay} = N_{events} / \epsilon \cdot {\cal L}
\label{limeqq} 
\ee
where, as in OPAL's analysis, the integrated luminosity is ${\cal L} = 10.3
\,pb^{-1}$, the signal detection efficiency\footnote{Our use of OPAL's 20\%
  signal efficiency is conservative.  OPAL considered pair-production of
  sequential leptons that decay via charged currents.  About one-tenth of the
  time, both W's decay leptonically; these $\ell\ell\nu\nu\nu\nu$ final
  states would be rejected by OPAL's cuts.  In considering cases where one or
  both of our heavy leptons decay via neutral currents, we have not included
  the analogous $\ell\ell\nu\nu\nu\nu$ events.  Thus a higher fraction of the
  events we did include should pass OPAL's cuts.}  is $\epsilon \approx
20\%$, and the number of (unseen) signal events is $N_{events} \approx 3$.
Thus an upper bound on the number of signal events implies an upper bound on
$\sigma_{production}\cdot B_{decay}$.  Inserting the branching fraction for
$\ell^H \ell^H$ pairs to visible final states, $B_{decay}$, as in equation
(\ref{Bdecay}) yields an upper bound on the production cross-section.  
Since we have already calculated the cross-section
($\sigma_{production}(\sqrt{s}=172 \ {\rm GeV})$) as a function of heavy
lepton mass, we can convert the bound on $\sigma_{production}$ into a
a $95\%$ c.l. lower bound on $m^H_\tau$:
\be
m^H_\tau > 79.8\ {\rm GeV} .
\ee
This is a great improvement over the bounds of order 40 GeV,
(\ref{mhtaub}) and (\ref{mhtaud}), we obtained earlier from precision
electroweak data in cases B and D where the tau is the only lepton to
have a weak-singlet partner.

Our new lower bound on $m^H_\tau$ further constrains the allowed
region of the $M_{\ell}{\,}$vs.${\,}v\lambda_{\ell}$ parameter space,
as illustrated in figure \ref{OPALlimfig}.  Contours on which the
heavy tau mass takes on the values $m_\tau^H=70,\, 79.8$ and $92$ GeV
are shown as a reference and to indicate how a tighter mass bound
would affect the size of the allowed region.

\begin{figure}[htb]
\rotatebox{90}{\resizebox{8cm}{14cm}{\includegraphics{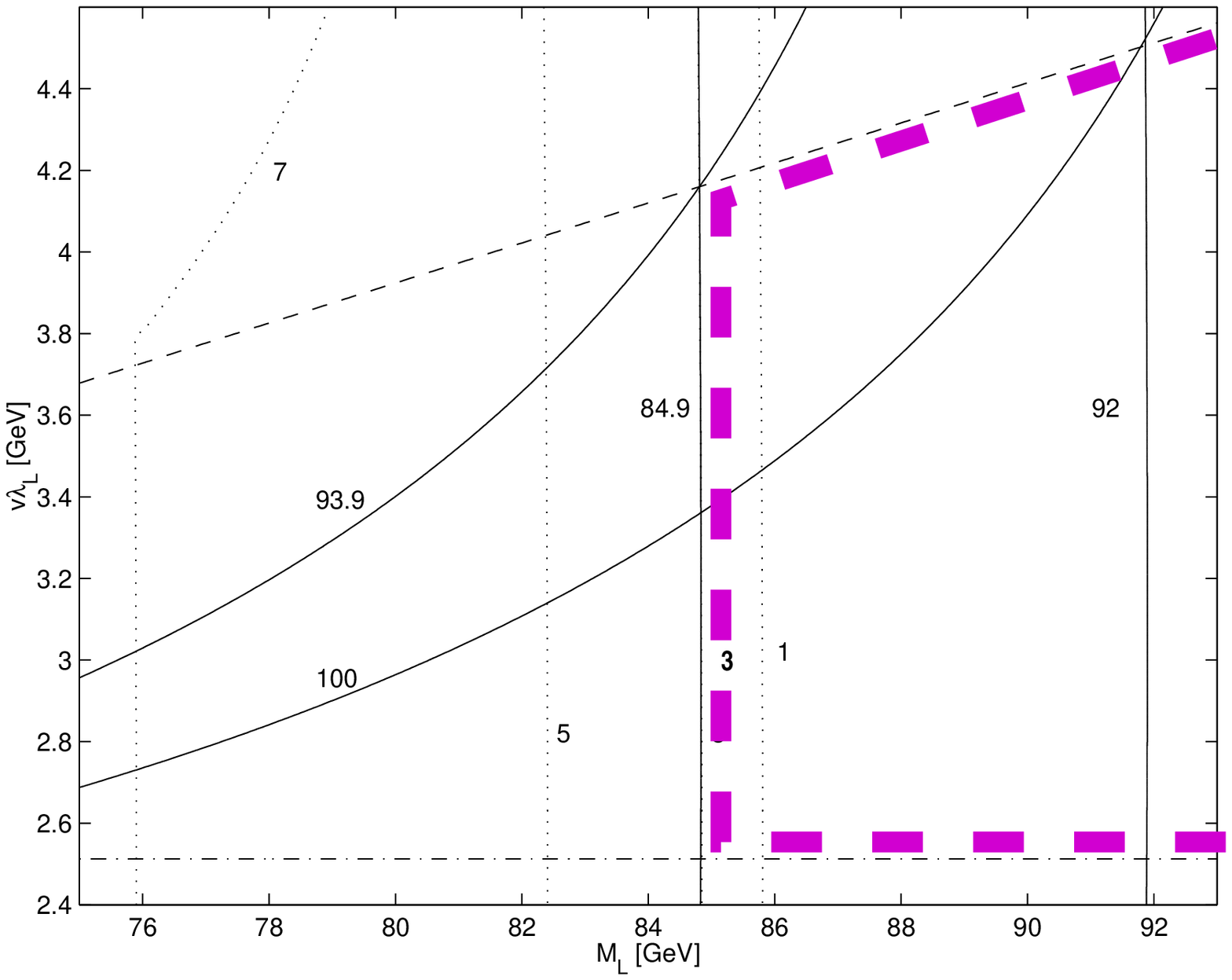}}}
\caption[OPALlim1]{\small Lower bounds on heavy lepton mass in models with three
  flavors of heavy lepton (Case A).    The heavy dashed curve encloses
  the region of the plot allowed by the combination of electroweak and direct
  search data.  The limits on $\sin{\phi_{\tau}}^2$ (dashed lines) and
  $v\lambda_{\ell}=m^L_{\tau}\sqrt{2}$ (dot-dash line) are as in Figure 1.
  Solid lines are
  contours of constant heavy lepton mass - those for $m^H_{e, \mu}$ are
  vertical, those for $m^H_\tau$ are curved.  Dotted lines are contours of
  constant number of OPAL signal events; the cusp shows where the
  pair-production threshold for $\tau^H$ is crossed. The mass limit for $e^H,
  \mu^H$ comes from the overlap of the $N_{events}=3$ dotted line with the
  $m^H_{e, \mu}=84.9 \ {\rm GeV}$ solid line; at these values of
  $v\lambda_\ell$ and $M_\ell$, the $\tau^H$ is above OPAL's pair-production
  threshold.  In calculating branching fractions, the scalar mass was set
  to 100 GeV.}
\label{OPALlim1fig}
\end{figure}

In case A, where $e$, $\mu$, and $\tau$ all have singlet
partners, the contributions from all three heavy leptons to the signal have
to be taken into account. While the $e^H$ and $\mu^H$ have nearly identical
masses and decays, the $\tau^H$ has slightly different properties.
By adding the contributions from all three flavors of heavy lepton, 
drawing the contour corresponding to $N_{events}=3$ on the
$M_{\ell}{\,}$vs.${\,}v\lambda_{\ell}$ parameter space, and comparing this
with contours of constant $m^H_\ell$ for each species, 
we obtain the 95\% c.l. lower bounds on all three heavy lepton
masses, as shown in figure \ref{OPALlim1fig}
\be
m^H_{e, \mu} > 84.9\ {\rm GeV}\ \ 
\ee
\be
m^H_\tau > 93.9\ {\rm GeV}\, .\label{consist2}
\ee
Note that the bound on $m^H_\tau$ comes simply from internal
consistency of the model (the values of $v\lambda_\ell$ and $M_\ell$
are flavor-universal), since it lies above OPAL's pair-production
threshold.  These bounds are a significant improvement over
those we obtained from precision data, i.e. (\ref{mhem}) and
(\ref{mhtaua}).

While calculating the lower limits on the $m^H_\ell$ required us to
assume a value for $M_\Phi$ (to evaluate $B_{decay}$), the result is
insensitive to the precise value chosen.  As noted in section 5.1, in the
allowed region of the $v\lambda_\ell$ vs. $M_\ell$ plane, LEP's lower bound
on the Higgs boson's mass applies to $\Phi$ so that min($m_\ell^H$) $\leq$
min($M_\Phi$).  In this case, $B(\ell^H \to \Phi \ell^L)$ is negligible.  

Our limits are also insensitive to the precise values of the small lepton
mixing angles $\slps_\ell$.  The production rate has little dependence on
$\slps_\ell$ because the $\ell^H\ell^H Z$ coupling (\ref{zcoupl}) is
dominated by the ``$-Q \sin^2\theta$'' term.  What little dependence there is on
$\slps_\ell$ decreases as $2 m^H_\ell$ approaches $\sqrt{s}$, and the mass
limits tend to be set quite close to the production threshold.  Moreover, the
branching fractions for the vector boson decays of the $\ell^H$ have only a
weak dependence on $\slps_\ell$.  Both the charged- and neutral-current decay
rates are proportional to $\slp_\ell$ (and the rate for decay via Higgs
emission is negligible), so that the mixing angle dependence in the branching
ratio comes only through factors of $\cos^2\phi_\ell$ which are nearly equal to
1.  As a result, our lower bounds on the heavy fermion masses will stand even
if improved electroweak measurements tighten constraints on the mixing
angles.

Because the mass limit tracks the pair-production threshold, stronger mass
limits can be set by data taken at higher center-of-mass energies.  Figure
\ref{rtsfig} shows $\sigma_{production}$ as a function of the heavy lepton
mass for several values of $\sqrt{s}$ and $\slp_\ell$.  As data from higher
energies provides a new, more stringent upper bound on $\sigma_{production}
\cdot B_{decay}$, one can read an improved lower bound on the heavy lepton
mass from figure \ref{rtsfig}. 

More generally, one can infer a lower mass limit on a heavy
mostly-weak-singlet lepton from other models using the same data by
inserting the appropriate factor of $B_{decay}$ in equation
(\ref{limeqq}). For models in which the mixing angles between ordinary
and singlet leptons are small and in which $B(\ell^H \to \Phi \ell^L)$
is small, our results apply directly.  This would be true, for
example, of some of the heavy leptons in the flavor-universal top
seesaw models \cite{burdev}.

\bigskip
Since the lower bound the LEP II data sets on the mass of the heavy
leptons is close to the kinematic threshold for pair production, it 
seems prudent to investigate whether single production
\be
e^+ e^- \to \ell^H \ell^L 
\ee
would give a stronger bound.  Single production proceeds only through $Z$
exchange (the $\gamma f^H f^L$ coupling is zero).  Moreover, 
equation (\ref{zcoupl}) shows that the $Z \ell^H \ell^L$ coupling is
suppressed by a factor of $\slps_\ell$; given the existing upper bounds on
the
mixing angles (\ref{casea})-(\ref{cased}), the suppression is by a factor
of at
least 10.  As a result, only a fraction of a single-production event is
predicted to have occurred (let alone have been detected) in the 10
pb$^{-1}$
of data each LEP detector has collected -- too little for setting a limit.

\subsection{Heavy quarks at the Tevatron}

New quarks decaying via mixing to an ordinary quark plus a heavy boson
would contribute to the dilepton events used by the Tevatron
experiments to measure the top quark production cross-section
\cite{D0-quark}\cite{CDF-quark}.  We will use the results of the
existing top quark analysis and see what additional physics is
excluded.  If evidence of new heavy fermions emerges in a future
experiment, it will be necessary to do a combined analysis that
includes both the top quark and the new fermions and that examines
multiple decay channels.

Here, we use the dilepton events observed at Run I to set limits on direct
production of new largely-weak-singlet quarks (our $q^H$).  These new quarks
are color triplets and would be produced with the same cross-section as
sequential quarks of identical mass.  However, their weak-singlet component
would allow the new states to decay via neutral-currents as well as
charged-currents.  This affects the branching fraction of the produced quarks
into the final states to which the experimental search is sensitive.

The D\O\ and CDF experiments searched for top quark events in the reaction
\be
p \bar{p} \to  Q \bar{Q} \to q W \bar{q} W \ \ \to\ \   
q \bar{q} \ell \nu_\ell \ell^\prime \nu_{\ell^\prime}\label{reactt} 
\ee
by selecting the final states with dileptons, missing energy, and at least
two jets.  Di-electron and di-muon events in which the dilepton invariant
mass was close to the Z mass were rejected in order to reduce Drell-Yan
background.  The top quark was assumed to have essentially 100\% branching
ratio to an ordinary quark (q) plus a W, as in the Standard Model.  The D\O\ 
(CDF) experiment observed 5 (9) dilepton events, as compared with 1.4 $\pm$
0.4 (2.4 $\pm$ 0.5) events expected from Standard 
model backgrounds and 4.1
$\pm$ 0.7 (4.4 $\pm$ 0.6) events expected from top quark production.  Thus,
D\O\ (CDF) measured the top production cross-section to be 5.5 $\pm$ 1.8 pb
($8.2 ^{+ 4.4}_{-3.4}$ pb).

In using this data to provide limits on the production of heavy quarks
in our models, we consider dilepton events arising from top quark
decays to be part of the background.  Hence, from D\O\ (CDF), we have
5 (9) dilepton events as compared with a background of 5.5 $\pm$ 0.8
(6.8 $\pm$ 0.8) events.  At 95\% confidence level, this implies an
upper limit of 5.8 (9.6) on the number of additional events that could
have been present due to production and decays of new heavy quarks.

How many $q^H$ would be produced and seen ?  The $q^H$ have the same QCD
production cross-section as a Standard Model quark of the same mass.  The
$q^H$ can decay by the same route as the top quark (\ref{reactt}).  About
10\% of the charged-current decays of pair-produced $q^H$ would yield final
states to which the FNAL dilepton searches were sensitive.  The neutral
current decays of the $q^H$ reduce the charged-current branching fraction
$B(q^H \to q^L W)$, but will not, themselves, contribute
significantly\footnote{A Higgs-like 
  scalar with a mass of order 130-150 GeV could have a relatively large
  branching fraction to two $W$ bosons \cite{hscaldcy} .   This might allow some
  neutral-current decays of $q^H$ to contribute to the dilepton sample and
  change our mass bounds slightly.}
to the dilepton sample since dileptons from $Z$ decays are specifically 
rejected and the $\Phi$ couplings to $e$ and $\mu$ are extremely small.
Then we estimate the fraction of heavy quark pair events that would
contribute to the dilepton sample as
\be
B_{decay}= B_{\ell\ell}\,(B_W)^2
\label{Bdecay1}
\ee
where $B_{\ell\ell}$ is the fraction of $W$ pairs in which both bosons
decay leptonically and $B_W \equiv B(q^H \to q^L W)$ is calculated in section
5.1 and shown in figure 5.

The number of dilepton events expected in a heavy-quark production experiment
with luminosity ${\cal L}$ and detection efficiency for dilepton events
$\epsilon$ is
\be
N^{q^H} = \sigma^{q^H} \cdot {\cal L} \cdot \epsilon \cdot B_{decay}
\label{chtheeq}
\ee
Similarly in top searches the total number of events is
\be
N^t = \sigma^t \cdot {\cal L} \cdot \epsilon \cdot B_{cc}
\cdot B_{\ell\ell}
\label{chtheeq1}
\ee
where $B_{cc}$ is the fraction of top quark pairs decaying via charged
currents.

In comparing the number of events expected for produced
top quarks with those for $q^H$ pairs, the values of $\epsilon$ and 
${\cal{L}}$ are the same; furthermore, $B_{cc}$ of equation
(\ref{chtheeq1})
is essentially 100\% . Therefore we may write
\be
\sigma^{q^H} \, (B_W)^2 = \sigma^{t} \ {N^{q^H}
  \over N^{t}} 
\label{chtheeq2}
\ee
Using the values which the CDF and D\O\ experiments have determined for the
three quantities on the right-hand side (cf. previous discussion), we find
\bea
\sigma^{q^H} \, (B_W)^2  &\leq& 7.8\ {\rm pb}\ \ ({\rm
  DO})\nonumber \\
&\leq& 12.0\ {\rm pb}\ \ ({\rm CDF}) .\label{chtheeq3}
\eea
The dilepton sample at the Tevatron is sensitive only to the presence
of the $d^H$ or $s^H$ quarks in our models.  The $u^H$, $c^H$ and
$t^H$ are, according to equations (\ref{mtoprev}), (\ref{mhtopa}) and
(\ref{mhuc}), too heavy to be produced, while the $b^H$ decay
dominantly by neutral instead of charged currents, due to Cabibbo
suppression.  Hence, this search tests only the models of case A, in
which the light ordinary fermions have weak-singlet partners.

\begin{figure}[ht]
\rotatebox{90}{\resizebox{8cm}{14cm}{\includegraphics{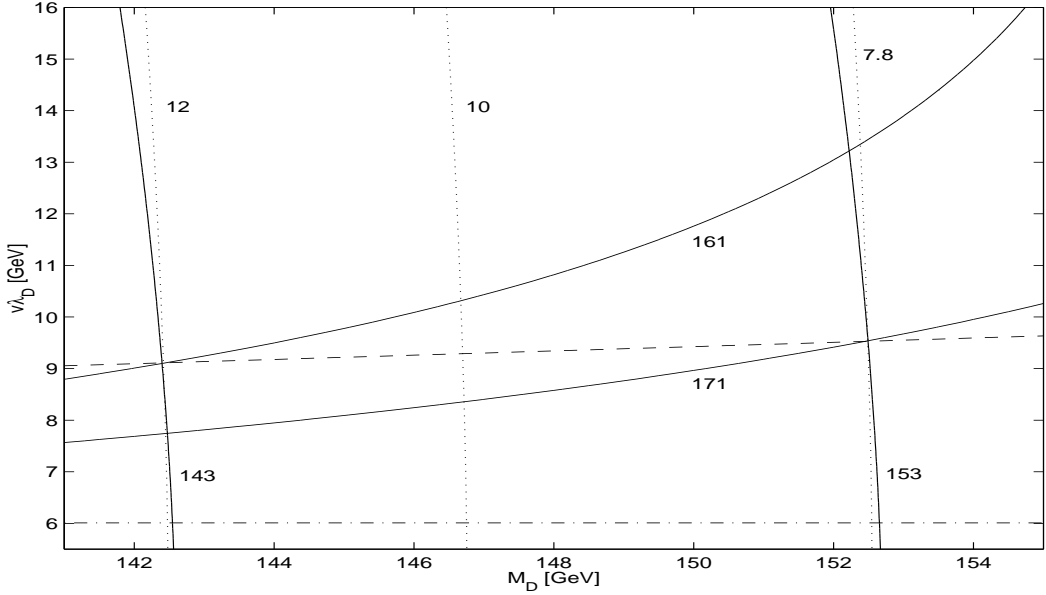}}}
\caption[TEVAlim]{\small A graphical representation of the Tevatron limits
  on heavy D quark masses.  The region allowed by precision electroweak tests
  (cf. fig 2) is bounded from above by the dashed line ($\sin{\phi_b}^2$) and
  from below by the dot-dashed line ($v\lambda_D$).  Dotted lines
  represent curves of constant $\sigma_{prod}\cdot (B_W)^2$ as
  in equation (\ref{chtheeq2}). Solid vertical lines are curves of
  constant $m^H_{d, s}$ ($m^H_b$); their overlap with the dotted lines
  defines the lower bound on the heavy fermion masses.  We set $M_\Phi=100$
  GeV in calculating branching fractions.}
\label{TEVAlimfig}
\end{figure}

Since the pair-production cross-section for $q^H$ is the same as that for a
heavy ordinary quark, we use the cross-section plots of reference
\cite{laenen} and our calculated branching fraction $B_{decay}$
(\ref{Bdecay1})  to translate equation (\ref{chtheeq3}) into
lower bounds on heavy fermion masses.  For Case A, in which both the $d^H$
and $s^H$ quarks can contribute to the dilepton sample, we find (with $M_\Phi
= 100$ GeV):
\bea
m^H_d, m^H_s &\geq& 153\ {\rm GeV}\ \ ({\rm DO})\nonumber \\
&\geq& 143\ {\rm GeV}\ \ ({\rm CDF}) \nonumber\\
 m^H_b &\geq& 171\ {\rm GeV}\ \ ({\rm DO})\nonumber \\
&\geq& 161\ {\rm GeV}\ \ ({\rm CDF}) .\label{STEVA2}
\eea
which are significantly stronger than those obtained from low-energy data in
section 4 and also stronger than the published limits on a fourth-generation
sequential quark \cite{pubbp}.  Note that since the $b^H$ decays almost
exclusively via neutral-currents due to Cabbibo suppression of the
charged-current mode, the lower bound on $m^H_b$ is, once again, an indirect
limit implied by internal consistency of the model.  In the scenarios where
only third-generation fermions have weak partners (Cases B and C), we can
obtain no limit on $m_b^H$.

More generally, one can use the same data to infer an upper limit on
the pair-production cross-section for heavy mostly-weak-singlet quarks
from other models by inserting the appropriate factor of $B_{W}$ in
equation (\ref{chtheeq3}).  After taking into account the
number of heavy quarks contributing, one can use the cross-section
vs. mass plots of \cite{laenen} to determine lower bounds for the
heavy quark masses.  For example, our cross-section limits
(\ref{chtheeq3}) apply directly to the heavy mostly-singlet quarks in
the dynamical top-seesaw models that are kinematically unable to decay
to scalars and decay primarily by charged-currents.  The corresponding
mass limit depends on how many such quarks are in the model.

\section{Oblique Corrections}
\label{sec:phen-flav}
\setcounter{equation}{0}

The presence of new singlet fermions present in our models will shift
the $S$ and $T$ parameters \cite{STU} from their Standard Model
values.  In this section, we evaluate these changes and explore the
limits they impose on the fermion masses and couplings and the mass
of the scalar, $\Phi$.  This analysis of one-loop oblique corrections
turns out to complement the analysis of tree-level effects on
precision data performed in section 3: the oblique corrections most
strongly limit the top quark mixing angle which the earlier analysis
could not directly constrain.

\begin{figure}[bt]
\rotatebox{90}{\resizebox{8cm}{14cm}{\includegraphics{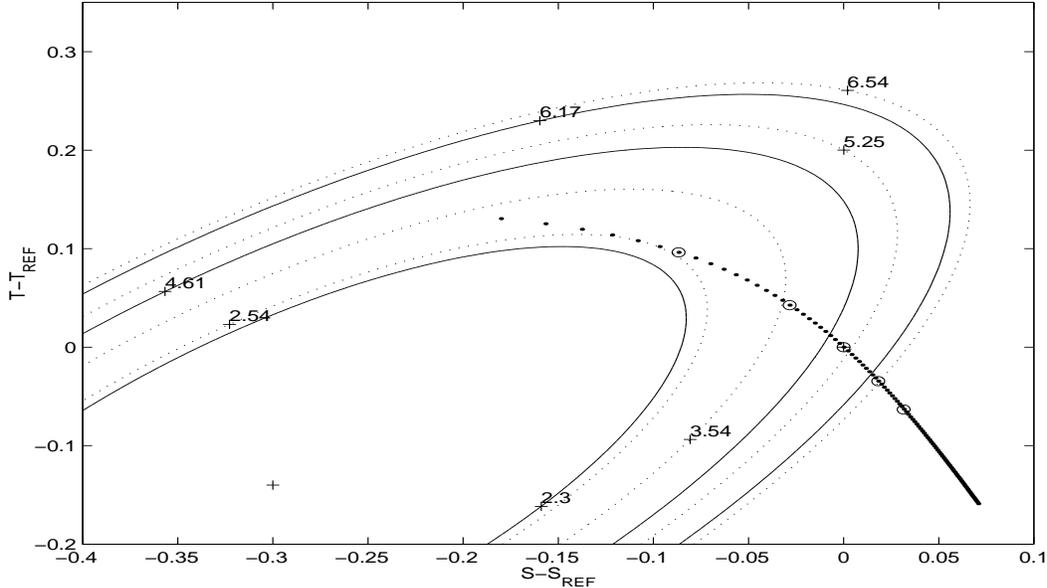}}}
\caption[stlim]{\small Comparing data on oblique corrections to
theoretical predictions.  Relative to the reference [$m_t = 173.9$
GeV, $m_H = 300$ GeV, $\alpha^{-1}(M_Z) = 128.9$], the cross shows the
best experimental fit to $S$ and $T$; the solid ellipses are at the
corresponding 68\%, 90\% and 95\% confidence levels \cite{take} for
two degrees of freedom.  The labels on both the solid and dotted
ellipses indicate $\Delta \chi^2$ relative to the experimental
best-fit point (cross).  The heavy dotted curve shows how the
predicted value of $S$ and $T$ in the Standard Model varies as the
scalar mass $m_\Phi$ is varied by steps of 10 GeV (see text); lower
masses are to the left.  The value of $m_\Phi$ corresponding to the
lowest $\chi^2$ (smallest dotted ellipse) is $\approx$ 80 GeV.}
\label{stlimfig}
\end{figure}

In calculating the values of $S$ and $T$ predicted by our models, we
started from the results of \cite{take}, which cite the experimental
values of $S$ and $T$ relative to the reference point [$m_t = 173.9$
GeV, $m_H = 300$ GeV, $\alpha^{-1}(M_Z) = 128.9$].  We included the
appropriately weighted variations of $m_t$ and $\alpha^{-1}$ and
obtained the minimal combined $\chi^2$ field on the $S-S_{ref}$ vs
$T-T_{ref}$ plane; we simultaneously obtained the corresponding
$m_t(S,T)$ and $\alpha^{-1}(S,T)$ that minimize $\chi^2$ for each pair
of S and T parameters. The minimal combined $\chi^2$ is presented in
in figure \ref{stlimfig}; the solid ellipses represent
joint 68.3\%, 90\%, and 95.4\% c.l. limits on S and T with variations
in $m_t$ and $\alpha^{-1}$ included.  Next, within the Standard Model
we allowed the Higgs mass to vary \cite{higgvary} from 40 GeV to 1 TeV
in steps of 10 GeV and obtained the ``best fit Higgs curve" shown in
figure \ref{stlimfig}; the circled points are at 100, 200, 300, 400,
and 500 GeV (smaller masses to the left).  The dotted ellipses in the
figure are contours of constant minimal combined $\chi^2$ whose
intersections with the ``best fit Higgs curve'' define the best fit
value and 68.3\%, 90\%, and 95.4\% c.l. limits on Higgs mass. These
values are respectively 80 GeV (in good agreement with
\cite{LEPEWWG}), 190 GeV, 310 GeV, and 400 GeV.
 
We then added the effects of the extra fermions on $S$ and $T$.  The
contribution of the singlet fermions to S was calculated numerically
using the formalism described in \cite{pbook}.  The contribution to
$T$ was found analytically \cite{CAR, CAR1} by summing the
vacuum-polarization diagrams containing the heavy and light
mass-eigenstate fermions present in the model of interest.  For
example, in models containing weak-singlet partners for only the $t$
and $b$ quarks, we find that the contribution of the $t^H$, $t^L$,
$b^H$ and $b^L$ states to the $T$ parameter is (in agreement with
\cite{them})
\bea
\alpha T \!&-&\!\alpha T_H = {3\,G_{F}
\over
{8\,\pi^2\,\sqrt{2}}}\bigg{[}{m_t^L}^2\,c_t^4+{m_b^L}^2\,c_b^4+{m_t^H}^2
\,s_t^4+{m_b^H}^2\,s_b^4 
\label{rho}\\
&-& 2\,{m_t^L}^4\,c_t^2\bigg{(}\,-\,{c_t^2 \over {m_t^L}^2}\,+\,{c_b^2
\over
  {{m_t^L}^2-{m_b^L}^2}}\,-\,{s_t^2 \over
{{m_t^L}^2-{m_t^H}^2}}\,+\,{s_b^2
  \over {{m_t^L}^2-{m_b^H}^2}}\bigg{)}\ln({m_t^L}^2)\nonumber\\
&-& 2\,{m_b^L}^4\,c_b^2\bigg{(}\,+\,{c_t^2 \over
{{m_b^L}^2-{m_t^L}^2}}\,-\,{c_b^2 \over
  {m_b^L}^2}\,+\,{s_t^2 \over {{m_b^L}^2-{m_t^H}^2}}\,-\,{s_b^2
  \over {{m_b^L}^2-{m_b^H}^2}}\bigg{)}\ln({m_b^L}^2)\nonumber\\
&-& 2\,{m_t^H}^4\,s_t^2\bigg{(}\,-\,{c_t^2 \over
{{m_t^H}^2-{m_t^L}^2}}\,+\,{c_b^2 \over
  {{m_t^H}^2-{m_b^L}^2}}\,-\,{s_t^2 \over {m_t^H}^2}\,+\,{s_b^2
  \over {{m_t^H}^2-{m_b^H}^2}}\bigg{)}\ln({m_t^H}^2)\nonumber\\
&-& 2\,{m_b^H}^4\,s_b^2\bigg{(}\,+\,{c_t^2 \over
{{m_b^H}^2-{m_t^L}^2}}\,-\,{c_b^2 \over
  {{m_b^H}^2-{m_b^L}^2}}\,+\,{s_t^2 \over
{{m_b^H}^2-{m_t^H}^2}}\,-\,{s_b^2
  \over {m_b^H}^2}\bigg{)}\ln({m_b^H}^2)\bigg{]}\nonumber
\eea   
where $T_H$ is the Higgs contribution, and $c_f$
($s_f$) is an abbreviation for $\cos\phi_f$ ($\sin\phi_f$).  To
isolate the extra contribution caused by the presence of the
weak-singlet partners for the $t$ and $b$ quarks, we must subtract off
the amount which $t$ and $b$ contribute in the
Standard Model \cite{CAR1}:
\be
\alpha T\,-\,\alpha T_H = {3\,G_{F}  
\over
{8\,\pi^2\,\sqrt{2}}}\bigg{[}{m_t^L}^2\,+\,{m_b^L}^2\,-\,{2{m_t^L}^2{m_b^L}^2
\over
{{m_t^L}^2-{m_b^L}^2}}\ln\bigg{(}{{m_t^L}^2
\over{m_b^L}^2}\bigg{)}\bigg{]}\label{rho1}
\ee
Note that (\ref{rho}) correctly reduces to (\ref{rho1}) in the limit
where singlet and ordinary fermions do not mix ($\slp_t, \slp_b \to
0$).  From the form of equation (\ref{rho}), we see that experimental
bounds on the magnitude of $T$ will constrain relatively heavy extra
fermions to have small mixing angles.

To illustrate how oblique effects constrain non-standard fermions, we
begin by including a weak-singlet partner only for the top quark; that
is, we send $\slp_b \to 0$ in equation (\ref{rho}).  For a given
scalar mass $m_\Phi$, we add to the Standard Model $S$ and $T$, the
additional contribution caused by mixing of an ordinary and
weak-singlet top quark.  For the $T$ parameter, this extra
contribution is the difference between expressions (\ref{rho}) and
(\ref{rho1}) with $\slp_b = 0$.  By construction, for
$s_t^2\rightarrow 0$ the new contributions to the S and T parameters
both go to zero (i.e. $\delta S=\delta T=0$).  When mixing is present
($s_t^2 \neq 0$), one has $\delta S < 0$ and $\delta T >0$, and the
predicted values of $S$ and $T$ lie above the ``best fit Higgs curve''

\begin{figure}
{\resizebox{14cm}{8cm}{\includegraphics{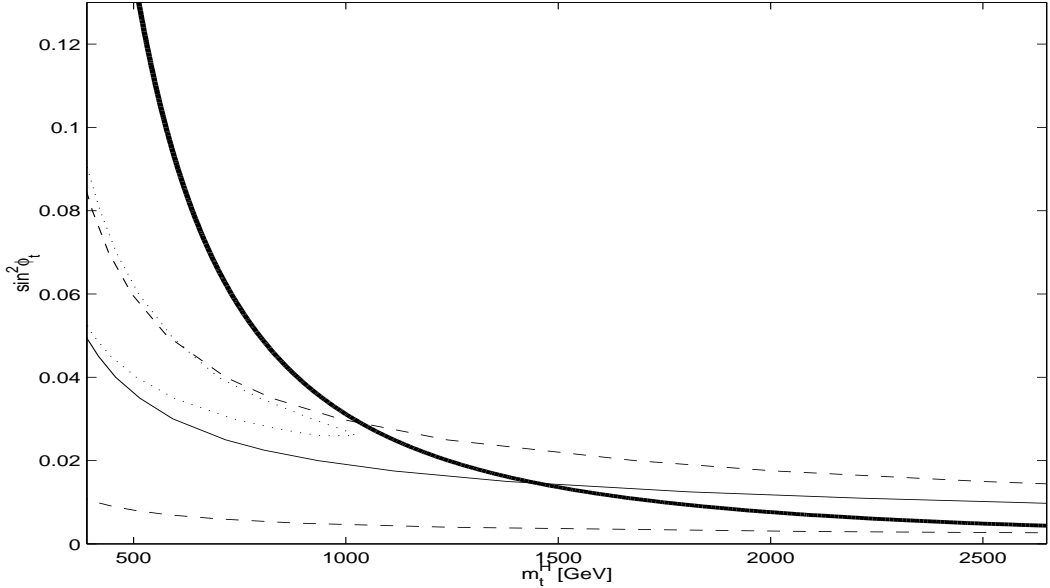}}}
\caption[mixtlim]{\small Lower bound on heavy top mass $m_t^H$ as a
function of heavy top mixing $\slp_t$.  Based on the oblique
corrections, for $m_\Phi$ = 100 GeV, the mass and mixing must fall
below the solid curve; for $m_\Phi$ = 350 GeV, they must fall in the
band between the dashed curves; for $m_\Phi$ = 520 GeV, they must lie
within the dotted curve.  The additional lower bound on $m_t^H$
\protect{\ref{mhtbcd}} is represented by the heavy solid curve; the
allowed region is to the right of this curve, leading to the
constraints (\protect{\ref{mhtlimm}) and (\protect{\ref{mhtlimm2}})}.}
\label{mixtlimfig}
\end{figure}

We deem ``allowed'' the values of $m_t^H$ and $\slp_t$ for which the
final values of $S$ and $T$ fall inside the dotted ellipse labeled
$\Delta \chi^2$ = 5.25 -- the 90\% c.l. ellipse for the Standard Model
alone.  In other words, we require that the model including new
physics agree with experiment at least as well as the Standard Model.
This allows us to trace out a region of allowed heavy top mass and
mixing for different values of $m_\Phi$, as illustrated in figure
\ref{mixtlimfig}.  Note that the presence of non-zero mixing of
ordinary and singlet top quarks enables a heavier scalar to be
consistent with the data\footnote{
For a discussion of related issues
for the Standard Model Higgs boson see \cite{tripre}.}.

As a complementary limit on $m_t^H$ and $\slp_t$, we note that the discussion in section 4
requires
\be
m^H_t  \ge  m^L_t \sqrt{1 + 1/\slp_t}.
\label{mhtbcd}
\ee
That is, for a given amount of mixing, the heavy top mass must
lie above some minimum value.  Combining these limits yields the
allowed region in the mixing vs. mass space in figure
\ref{mixtlimfig}.  For example, 
\bea
{\rm For}\ \  m_\phi = 100 {\rm GeV},\ \ \ \ m_t^H &\gae& 
1450{\rm GeV}\label{mhtlimm}\\
                                         \slp_t & \lae& .015 
\nonumber\\ 
{\rm For}\ \  m_\phi = 350 {\rm GeV},\ \ \ \ m_t^H &\gae& 
1040{\rm GeV}\label{mhtlimm2}\\
                                   \slp_t & \lae&  .031 \nonumber
\eea
As illustrated in figure \ref{mixtlimfig}, if the scalar's mass, $M_\Phi$, rises above 520
GeV, the regions of top mass and mixing allowed by oblique corrections
by equation (\ref{mhtbcd}) cease to intersect; this provides an upper
bound on the scalar mass.

\begin{figure}
\rotatebox{90}{\resizebox{8cm}{14cm}{\includegraphics{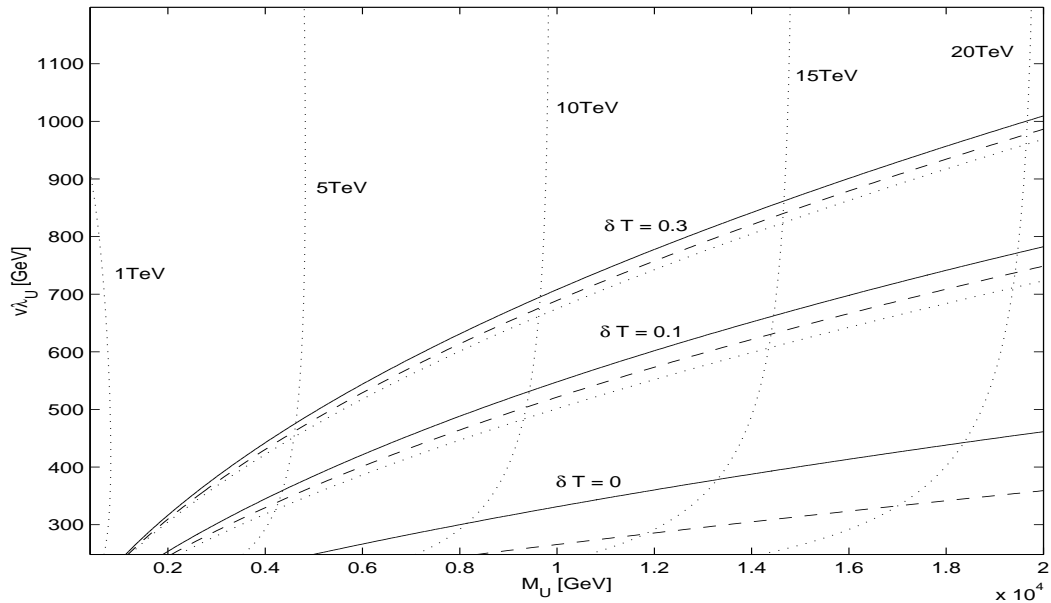}}}
\caption[rho3lim]{\small The effect of b-quark mixing on $T$.
Contours of constant $\delta T$ are shown in the up-sector
$M_U\,$v.s.$\,v\lambda_U$ parameter space.  Three representative
values ($\delta T$ = 0, 0.1, 0.3) are shown for several values of
heavy b-quark mass and mixing.  Solid curves correspond to $m^H_b=5$TeV
and $\sin^2\phi_b=0.00090$ (equivalently, $M_D=5$TeV,
$v\lambda_D=150$GeV). Dashed curves are $m^H_b=5$TeV and
$\sin^2\phi_b=0.00040$ ($M_D=5$TeV, $v\lambda_D=100$GeV). Dotted
curves are for $m^H_b=0.55$TeV and $\sin^2\phi_b=0.00027$
($M_D=0.5$TeV, $v\lambda_D=10$GeV).  The nearly-vertical dotted curves
of constant $m_t^H$ are shown for reference.}
\label{rho3limfig}
\end{figure}

To apply oblique-correction constraints to our models, we need to
include weak-singlet partners for quarks other than the top quark.
Since these fermions contribute little to $S$ \cite{STU}, we can
illustrate the effects of including other singlet fermions by showing
how they affect the $T$ parameter.  First, we
include the singlet partner for the $b$ quark, as in equation
(\ref{rho}).  We can interpret the result using figure
\ref{rho3limfig}, which shows the value of $T$ within the
coupling-mass plane for the up-sector quarks.  For reference, dotted
nearly-vertical curves of constant heavy top mass $m_t^H$ are shown.
The main contents of the figure are the three sets of curves labeled
$\delta T$ = [0.3, 0.1, 0], where $\delta T$ is the contribution due
to mixing between ordinary and singlet fermions.  Within each set, the
separate curves correspond to different values of the heavy b mass and
mixing.  The solid curve obtains for $m_b^H = 5$ TeV and $\slp_b =
0.00090$; the dashed curve, for $m_b^H = 5$ TeV and $\slp_b =
0.00040$; the doted curve, for $m_b^H = 0.55$ TeV and $\slp_b =
0.00027$.  Looking at the region where $m_t^H$ is of order a few TeV,
we see that the influence of the b-quark is small.  Including the
effects of partners for the other fermions yields a generalized
version of equation \ref{rho} and similar results. Thus the lower
bounds on $m_t^H$ we found earlier by considering only mixing for the
top quark will not be much altered by including mixing for the other
quarks, as in our models A, B, and C.

\section{Conclusions}
\label{sec:concl}

Precision electroweak data constrains the mixing between the ordinary
standard model fermions and new weak-singlet states to be small; our
global fit to current data provides upper bounds on those mixing
angles.  Even when the mixing angles are small, it is possible for
most of the exotic mass eigenstates which are largely weak-singlets to
be light enough to be accessible to collider searches for new
fermions.  We have analyzed in detail a class of models in which
flavor-symmetry breaking is conveyed to the ordinary fermions by soft
symmetry-breaking mass terms connecting them to new weak-singlet
fermions; such models have a natural GIM mechanism and a flavor
structure that is stable under renormalization.  By calculating the
branching rates for the decays of the heavy mass-eigenstates (which
are significantly influenced by their being primarily weak-singlet in
nature) we have been able to adapt results from searches for new
sequential fermions to further constrain our models.  We find that
direct searches at LEP II now imply that the heavy leptons $\ell^H$
must have masses in excess of 80-90 GeV; those limits are not
sensitive to the precise values of the small mixing angles.  Current
Tevatron data indicates that heavy quark states $d^H$ and $s^H$ could
be as light as about 140-150 GeV, while the mostly-weak-singlet $b^H$
must weigh at least 160-170 GeV.  In addition, the new fermions'
contributions to the oblique corrections allow the scalar $\Phi$ to
have a relatively large mass (up to about 500 GeV) while remaining
consistent with the data.  Oblique corrections also constrain the
mixing and mass of the the heavy top state which is mostly
weak-singlet; in particular, $m_t^H$ must be at least 1 TeV.  Finally,
we have indicated how our phenomenological results may be generalized
to related models, including the dynamical top-seesaw theories.

\bigskip
\appendix
\section{Appendix: Mixing effects on electroweak observables}
\label{sec:appx}
\setcounter{equation}{0}

This appendix contains the expressions for the leading-order (in mixing
angles) changes to electroweak observables in the presence of fermion mixing.
The expressions were derived using equations (\ref{lwco} -- \ref{zcoupl}) and
the general approach of reference \cite{BGKLM}.

\bea
\Delta \Gamma_Z / \Gamma^{SM}_Z &=& 0.603\,(\slp_e + \slp_\mu)  
         - 0.072\,\slp_\tau \\
         &-& 0.3535\,(\slp_d + \slp_s + \slp_b) 
         - 0.287\,(\slp_u + \slp_c) \nonumber
\eea
\bea
\Delta \sigma_h / \sigma^{SM}_h &=& -1.409\,\slp_e 
         + 0.736\,\slp_\mu 
         + 0.072\,\slp_\tau \\ 
         &-& 0.1515\,(\slp_d + \slp_s + \slp_b) 
         - 0.124\,(\slp_u + \slp_c)\nonumber
\eea
\be
\Delta A_{\tau}(P_\tau)\, =\, \Delta A_{e}(P_\tau)\, =\, \Delta A_{LR}\, =\,
-0.5180\, \slp_e + 1.2870\, \slp_\mu 
\ee
\bea
\Delta R_b / R^{SM}_b &=& -0.0295 \,(\slp_e + \slp_\mu)  
         + 0.505\,(\slp_d + \slp_s) \nonumber \\
         &-& 1.78\,\slp_b 
         + 0.411\,(\slp_u + \slp_c)
\eea
\bea
\Delta R_c / R^{SM}_c &=& 0.0605\,(\slp_e + \slp_\mu)  
         + 0.505\,(\slp_d + \slp_s + \slp_b) \nonumber \\
         &+& 0.411\,\slp_u 
         - 1.999\,\slp_c 
\eea
\be
\Delta A_{FB}^b = -0.3300\,\slp_e + 0.8500\,\slp_\mu - 0.0161\,\slp_b  
\ee
\be
\Delta A_{FB}^c = -0.1785\,\slp_e + 0.6665\,\slp_\mu - 0.0875\,\slp_c  
\ee
\be
\Delta {\cal{A}}_b  = 0.1052\, (\slp_e + \slp_\mu) - 0.1472\, \slp_b
\ee
\be
\Delta {\cal{A}}_c  = 0.5719\, (\slp_e + \slp_\mu) - 0.7997\, \slp_c
\ee
\be
\Delta Q_W(Cs)  = 72.7663\,\slp_e - 0.7239\,\slp_\mu + 211.0024\,\slp_d
- 187.9988\,\slp_u
\ee
\be
\Delta Q_W(Tl)  = 111.396\,\slp_e - 4.920\,\slp_\mu + 327\,\slp_d
- 285\,\slp_u
\ee
\bea
\Delta R_e / R^{SM}_e &=& 2.275 \,\slp_e  
         + 0.130\,\slp_\mu  
         - 0.505\,(\slp_d + \slp_s + \slp_b) \nonumber \\
        &-& 0.411\,(\slp_u + \slp_c)
\eea
\bea
\Delta R_\mu / R^{SM}_\mu &=& 0.130 \,\slp_e  
         + 2.275 \,\slp_\mu  
         - 0.505\,(\slp_d + \slp_s + \slp_b) \nonumber\\ 
         &-& 0.411\,(\slp_u + \slp_c)
\eea
\bea
\Delta R_\tau / R^{SM}_\tau &=&  0.130\,(\slp_e + \slp_\mu) 
         + 2.145\,\slp_\tau  \\
         &-& 0.505\,(\slp_d + \slp_s + \slp_b) 
         - 0.411\,(\slp_u + \slp_c)\nonumber
\eea
\be
\Delta A_{FB}^e = -0.1230\,\slp_e + 0.3070\,\slp_\mu  
\ee
\be
\Delta A_{FB}^\mu = 0.0920\,(\slp_e +\slp_\mu)  
\ee
\be
\Delta A_{FB}^\tau = 0.0920\,\slp_e + 0.3070\,\slp_\mu - 0.2150\,\slp_\tau  
\ee
\be
\Delta A_{FB}^s = -0.3300\,\slp_e + 0.8500\,\slp_\mu - 0.0161\,\slp_s  
\ee
\be
\Delta M_W / M^{SM}_W = 0.1065\,(\slp_e + \slp_\mu)  
\ee
\be
\Delta g_{eV}(\nu e \rightarrow \nu e) = 0.1720\, \slp_e - 0.3650\,
\slp_\mu
\ee
\be
\Delta g_{eA}(\nu e \rightarrow \nu e) = 0.5000\, \slp_e - 0.5060\,
\slp_\mu
\ee
\be
\Delta g_L^2(\nu N \rightarrow \nu X) = 0.1220\, \slp_e + 0.7260\,
\slp_\mu -
0.4280\, \slp_d - 0.3445\, \slp_u
\ee
\be
\Delta g_R^2(\nu N \rightarrow \nu X) = -0.0425\, \slp_e + 0.0179\,
\slp_\mu
\ee
\be
\Delta R_{\pi} / R^{SM}_{\pi} = - \slp_e + \slp_\mu \qquad {\rm where}\
R_\pi 
\equiv {\Gamma(\pi \to e \bar\nu_e) \over
          {\Gamma(\pi \to \mu \bar\nu_\mu)}}
\ee
\be
\Delta R_{e\tau} / R^{SM}_{\tau} = \slp_\mu - \slp_\tau \qquad {\rm where}\ 
\ \ R_{e\tau} \equiv {\Gamma(\tau \to e \bar\nu_e \nu_\tau) \over 
          {\Gamma(\mu \to e \bar\nu_e \nu_\mu)}}
\ee
\be
\Delta R_{\mu\tau} / R^{SM}_{\mu\tau} = \slp_e - \slp_\tau \qquad {\rm
where}\ 
\ \ R_{\mu\tau} \equiv {\Gamma(\tau \to \mu \bar\nu_\mu \nu_\tau) \over 
          {\Gamma(\mu \to e \bar\nu_e \nu_\mu)}}
\ee

\bigskip
\section{Appendix: Details of heavy fermion decays}
\label{sec:appx2}
\setcounter{equation}{0}

This appendix contains details relevant to the heavy fermion decays
discussed in section 5.1.

At tree-level and neglecting final-state light fermion masses, the kinematic
factors F(x,y) and G(x,y) referred to in the text have the following form:
\bea
F(x,y)&=&{1 \over x^2}\{2\,x\,(2-x)+[3\,(x-1)+y^2]A(x,y)\label{ffun}\\
&+&[(x-1)^2(x+2)+3\,y^2(x-2)]B(x,y)\}\nonumber 
\eea
\bea
G(x,y)&=&{1 \over x^2}\{x\,(4x-3)+[x\,(4-x)-3+y^2]A(x,y)\label{gfun}\\
&+& 2[(x-1)^2+y^2(2x-3)]B(x,y)\}\nonumber
\eea
where A and B are given by
\bea
A(x,y)=\ln\bigg{[}1-{ x\,(x-2) \over {(x-1)^2+y^2}}\bigg{]}\label{afun}\\
B(x,y)={1 \over y}\,\bigg{[}\tan^{-1}\bigg{(}{1 \over
  y}\bigg{)}-\tan^{-1}\bigg{(}{{1-x} \over y}\bigg{)}\bigg{]}\label{bfun}
\eea

To check our general expressions for the decay rates, we evaluated their
behavior in the limiting cases where the decaying heavy fermion is either
much more massive or much less massive than the vector or scalar boson
involved in its decay.
Equations \ref{decayyZ/W} and \ref{ffun} for vector-boson decays yield
asymptotic behavior
\be
\!\Gamma(f_i^H \rightarrow f_j^L V)\, \stackrel{\scriptstyle
  m^{\scriptscriptstyle H}_{f_i} \gg M_{\scriptscriptstyle
    V}}{\longrightarrow}\,{(c^V_{ij})^2 \over {32
    \,\pi}}\,{{m^H_{f_i}}^3 \over {M_V}^2}\bigg{(}1-{{M_V}^2 \over
{m^H_{f_i}}^2}\bigg{)}^2\, 
\bigg{(}1+2{{M_V}^2 \over {m^H_{f_i}}^2}\bigg{)}\label{asdecayZ/W} 
\ee
\bea
\Gamma(f_i^H \rightarrow f_j^L V \rightarrow f_j^L f_k^L
f_l^L)\,\stackrel{\scriptstyle
  m^{\scriptscriptstyle H}_{f_i} \ll M_{\scriptscriptstyle
    V}}{\longrightarrow}&\,&\label{as1decayZ/W}\\
{(c^V_{ij}\,c^V_{kl})^2
  \over {3\,\pi^3\, 2^9}}{{m^H_{f_i}}^5 \over
  M_V^4}\,\bigg{(}1\!&+&\!{3 \over
  5}\,{{m^H_{f_i}}^2 \over M_V^2}\,+\,{2
  \over 5}\,{{m^H_{f_i}}^4 \over M_V^4}\,+\,{2 \over 7}\,{{m^H_{f_i}}^6
\over
  M_V^6}\,+\cdots\bigg{)}\nonumber
\eea
where V may be either Z or W. Equations \ref{decayH} and \ref{gfun} for
scalar-boson decays yield
\be
\Gamma(f_i^H \rightarrow f_j^L \Phi)\, \stackrel{\scriptstyle
  m^{\scriptscriptstyle H}_{f_i} \gg M_{\scriptscriptstyle
\Phi}}{\longrightarrow}\,{(c^H_{ij})^2 \over {32\,
    \pi}}\,{{m^H_{f_i}}^3 \over M_\Phi^2}\bigg{(}1-{M_\Phi^2 \over
{m^H_{f_i}}^2}\bigg{)}^2\label{asdecayH}\\
\ee
\bea
\Gamma(f_i^H \rightarrow f_j^L \Phi \rightarrow f_j^L f_k^L
f_l^L)\,\stackrel{\scriptstyle
  m^{\scriptscriptstyle H}_{f_i} \ll M_{\scriptscriptstyle
\Phi}}{\longrightarrow}&\,&\label{as1decayH}\\
{(c^H_{ij}\,c^H_{kl})^2
  \over {3\,\pi^3 \, 2^{11}}}{{m^H_{f_i}}^5 \over
  M_\Phi^4}\,\bigg{(}1&+&{4 \over 5}\,{{m^H_{f_i}}^2
  \over M_\Phi^2}\,+\,{3 \over 5}\,{{m^H_{f_i}}^4 \over M_\Phi^4}\,+\,{16 \over
  35}\,{{m^H_{f_i}}^6 \over M_\Phi^6}\,+\cdots\bigg{)} \ . \nonumber
\eea

\newpage

\bigskip
\centerline{\bf Acknowledgments}
\vspace{12pt}

The authors thank R.S. Chivukula, G. Burdman, C. Hoelbling, K. Lane,
and K. Lynch for useful conversations.  E.H.S. thanks the Aspen Center
for Physics for hospitality during the completion of a portion of the
work.  E.H.S. acknowledges the support of an NSF Faculty Early Career
Development (CAREER) award and an DOE Outstanding Junior Investigator
award.  {\em This work was supported in part by the National Science
Foundation under grant PHY-9501249, and by the Department of Energy
under grant DE-FG02-91ER40676.}


\end{document}